\documentstyle[12pt,psfig,graphicx]{article} 
\makeatletter \def\@cite#1#2{{[{#1}]\if@tempswa\typeout {IJCGA
warning: optional citation argument ignored: `#2'} \fi}}

\newcount\@tempcntc
\def\@citex[#1]#2{\if@filesw\immediate\write\@auxout{\string\citation{#2}}\fi
  \@tempcnta\z@\@tempcntb\m@ne\def\@citea{}\@cite{\@for\@citeb:=#2\do
    {\@ifundefined
       {b@\@citeb}{\@citeo\@tempcntb\m@ne\@citea\def\@citea{,}{\bf ?}\@warning
       {Citation `\@citeb' on page \thepage \space undefined}}%
    {\setbox\z@\hbox{\global\@tempcntc0\csname b@\@citeb\endcsname\relax}%
     \ifnum\@tempcntc=\z@ \@citeo\@tempcntb\m@ne
       \@citea\def\@citea{,}\hbox{\csname b@\@citeb\endcsname}%
     \else
      \advance\@tempcntb\@ne
      \ifnum\@tempcntb=\@tempcntc
      \else\advance\@tempcntb\m@ne\@citeo
      \@tempcnta\@tempcntc\@tempcntb\@tempcntc\fi\fi}}\@citeo}{#1}}
\def\@citeo{\ifnum\@tempcnta>\@tempcntb\else\@citea\def\@citea{,}%
  \ifnum\@tempcnta=\@tempcntb\the\@tempcnta\else
   {\advance\@tempcnta\@ne\ifnum\@tempcnta=\@tempcntb \else 
\def\@citea{--}\fi
    \advance\@tempcnta\m@ne\the\@tempcnta\@citea\the\@tempcntb}\fi\fi}
\makeatother

\def\boxit#1{\leavevmode\thinspace\hbox{\vrule\vtop{\vbox{\hrule%
        \vskip3pt\kern1pt\hbox{\vphantom{\bf/}\thinspace\thinspace%
        {\bf#1}\thinspace\thinspace}}\kern1pt\vskip3pt\hrule}\vrule}%
        \thinspace}
\def\Boxit#1{\noindent\vbox{\hrule\hbox{\vrule\kern3pt\vbox{
\advance\hsize-7pt\vskip-\parskip\kern3pt\bf#1 \hbox{\vrule height0pt
depth\dp\strutbox width0pt} \kern3pt}\kern3pt\vrule}\hrule}}

\newcommand{\Hh}{\lower1.2ex\hbox{$\stackrel{\textstyle
H}{\footnotesize\sim}$}}
\newcommand{\Hho}{\lower1.2ex\hbox{$\stackrel{\textstyle
H_1}{\footnotesize\sim}$}}
\newcommand{\Hhw}{\lower1.2ex\hbox{$\stackrel{\textstyle
H_2}{\footnotesize\sim}$}}
\newcommand{\h}{\lower1.2ex\hbox{$\stackrel{\textstyle
h}{\footnotesize\sim}$}}

\newcommand{\gsim}{\lower.7ex\hbox{$\;\stackrel{\textstyle>}{\sim}\;$}}
\newcommand{\lsim}{\lower.7ex\hbox{$\;\stackrel{\textstyle<}{\sim}\;$}}
\newcommand{\be}{\begin{equation}} \newcommand{\ee}{\end{equation}}
\newcommand{\beq}{\begin{equation}} \newcommand{\eeq}{\end{equation}}
\newcommand{\bea}{\begin{eqnarray}} \newcommand{\eea}{\end{eqnarray}}

\def\baselinestretch{1}
\begin{document}
\catcode`@=11 \newtoks\@stequation
\def\subequations{\refstepcounter{equation}%
\edef\@savedequation{\the\c@equation}%
\@stequation=\expandafter{\theequation}
\edef\@savedtheequation{\the\@stequation}
\edef\oldtheequation{\theequation}
\def\theequation{\oldtheequation\alph{equation}}}
\def\endsubequations{\setcounter{equation}{\@savedequation}%
\@stequation=\expandafter{\@savedtheequation}%
\edef\theequation{\the\@stequation}\global\@ignoretrue

\noindent} \catcode`@=12
\begin{titlepage}

\title{{\bf  Hints on the high-energy seesaw mechanism from the 
low-energy neutrino spectrum
}} 
\vskip3in \author{{\bf J.A. Casas$^1$}, {\bf A. Ibarra$^2$} and
{\bf F. Jim\'enez-Alburquerque$^1$\footnote{\baselineskip=16pt {\small E-mail addresses: {\tt
alberto.casas@uam.es, alejandro.ibarra@desy.de, fernando.jimenez@uam.es}}}}
\hspace{3cm}\\
{\small $^1$IFT-UAM/CSIC, C-XVI, Univ. Aut\'onoma de Madrid, 28049 Madrid, Spain}\\
{\small $^2$DESY,  Theory Group, Notkestrasse 85, D-22603 Hamburg, Germany}.
}  \date{}  \maketitle  \def\baselinestretch{1.15}
\begin{abstract}
\noindent
It is an experimental fact that the mass ratio for the two 
heavier neutrinos, $h=m_3/m_2\lsim 6$, is much smaller than the typical quark and lepton
hierarchies, which are 
${\cal O}(20-300)$.
We have explored whether this peculiar pattern of
neutrino masses can be a consequence of the peculiar way they are generated
through a see-saw mechanism, determining
1) How the present experimental data restrict the structure
of the high-energy seesaw parameters and 2) Which choices, among
the allowed ones, produce more naturally
the observed pattern of neutrino masses.
We have studied in particular  if starting with hierarchical neutrino
Yukawa couplings, as for the other fermions, one can naturally get
the observed $h\lsim 6$ ratio. 
To perform the analysis we have put forward a top-down parametrization
of the see-saw mechanism in terms of (high-energy) basis-independent quantities.
Among the main results, we find that in most cases $m_2/m_1\gg m_3/m_2$, so 
$m_1$ should be extremely tiny.
Also, the $V_R$ matrix
associated to the neutrino Yukawa couplings has a far from random structure,
naturally resembling $V_{\rm CKM}$. In fact we show that identifying $V_R$ 
and $V_{\rm CKM}$, as well as neutrino and $u-$quark Yukawa couplings can
reproduce $h^{\rm exp}$ in a highly non-trivial way, which is very suggestive.
The physical implications of these results are also discussed.

\end{abstract}

\thispagestyle{empty}
\vspace*{0.2cm} \leftline{December 2006} \leftline{}

\vskip-22cm \rightline{IFT-UAM/CSIC-06-55}
\rightline{DESY 06-232} 
\rightline{hep-ph/0612289} \vskip3in

\end{titlepage}
\setcounter{footnote}{0} \setcounter{page}{1}
\newpage
\baselineskip=20pt

\noindent

\section{Introduction}
The flavour structure of the leptonic sector of the Standard Model shows 
challenging differences with respect to the hadronic one. Much attention
has been attracted by the neutrino mixing matrix, $U_{\rm MNS}$, which
presents two large mixing angles and a small one, in contrast to the 
three small mixing angles of the CKM matrix. On the other hand, the neutrino 
spectrum is not as well known as the neutrino mixings. In particular,
we still do not know whether the spectrum has a normal or an inverse
hierarchy (i.e. whether the most split neutrino is the heaviest or the
lightest), or whether it is quasi-degenerate \cite{Gonzalez-Garcia:2002dz}. 
However, the amount of
available information allows us to notice that, in either case, the 
pattern of neutrino masses is neatly different from those of quarks 
and charged-leptons. According to the last analyses of 
neutrino oscillation experiments \cite{Maltoni:2004ei}, the two independent $\nu-$mass
splittings are (at 2$\sigma$)
\bea
\label{masssplits}
\Delta m^2_{\rm sol}= (7.3-8.5) \times 10^{-5}\ \ {\rm eV}^2,\;\;\;\;\; 
\Delta m^2_{\rm atm}= (2.2-3.0) \times 10^{-3}\ \ {\rm eV}^2.
\eea
Hence, even in the case of a normal hierarchy, the mass of the heaviest neutrino
is at most $\sim 6$ times the mass of the second heaviest one. 
The precise value depends on the mass of the lightest neutrino, as shown in Fig.~1.
%
\begin{figure}
\centerline{\vbox{
\psfig{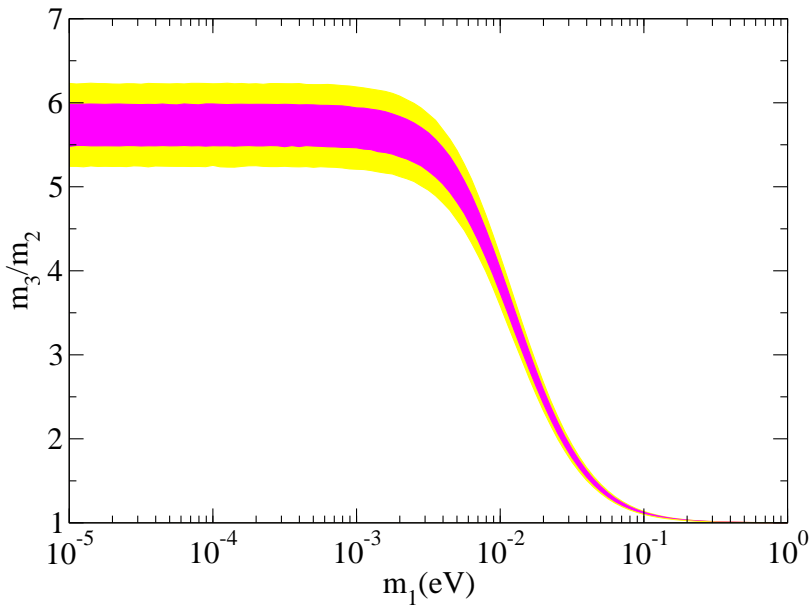}}}
\caption
{\footnotesize  
Experimental mass ratio of the two heavier neutrinos, $m_2/m_3$, vs. the mass of the lightest, $m_1$, in the case of normal hierarchy.}
\end{figure}
%
%
This contrasts
to the hierarchy observed in quarks and charged leptons, where the 
typical mass ratios 
are ${\cal O}(20)$ (for $d-$quarks and $\mu$/$\tau$ leptons) 
and ${\cal O}(300)$ (for $u-$quarks and $e$/$\mu$ leptons) \cite{Yao:2006px}.
Of course, if the $\nu-$spectrum is
quasi degenerate or with inverted hierarchy, the difference with the 
mass pattern of the other fermions is much more conspicuous. In any case
we can safely conclude that the hierarchy between the two heaviest neutrinos
is much softer than the one for the corresponding quarks or charged leptons.

According to the see-saw mechanism \cite{seesaw}, which is the
most popular mechanism for generating neutrino masses,
these arise in a slightly more complicated way than the masses of quarks and 
charged leptons. Namely, beside the conventional Yukawa couplings
between the Higgs, the left--handed and the right--handed neutrinos, 
one assumes Majorana masses for the right--handed
ones. Upon decoupling of the latter, the light neutrino
states have an effective Majorana mass matrix, 
${\cal M}_\nu\propto {\bf Y}^T {\bf M}^{-1}{\bf Y}$,
where ${\bf Y}$ is the initial matrix of Yukawa couplings and
${\bf M}$ is the Majorana mass matrix of the right-handed neutrinos.
So, unlike quarks and charged leptons, neutrino masses are not
proportional to the Yukawa couplings.
Then one may wonder whether the peculiar pattern of
neutrino masses can be a consequence of the peculiar way they are generated.
If so, the spectrum of light neutrinos may shed light on the unknown
features of the seesaw mechanism. In particular one may
ask 1) How the present experimental data restrict the structure
of the high-energy seesaw parameters and 2) Which choices, among
the allowed ones, produce more naturally
(i.e. without unpleasant fine-tunings) the observed pattern of neutrino masses.
In other words, one can examine how possible and how plausible is for the seesaw
mechanism to reproduce the experimental data, and what is the corresponding 
information that we can learn about the underlying high-energy theory.
Also, from such analysis, one can hopefully extract hints on the still
unknown part of the low-energy $\nu-$spectrum. 
The investigation of these questions and their physical implications is 
the goal of this paper.

In sect.~2 we fix the notation and put forward a basis-independent top-down parametrization for the see-saw, which is specially useful to study the pattern of $\nu-$masses. 
We discuss how the $V_R$ mixing matrix associated to ${\bf Y}$ plays here a key role.
In sect.~3 we analyze the 2-neutrino case, as a simple and useful warm-up. In sect.~4 we study the 3-neutrino case. We give general analytical results, completing (and confirming) them with numerical surveys. We pay special attention to the possibility that the $\nu-$spectrum
could arise from hierarchical Yukawa couplings, as for the other fermions, and work out
the required structure of the high energy parameters and some consequences for the unknown
part of the low-energy $\nu-$spectrum. In sect.~5 we explore suggestive ans\"atze for the
$V_R$, showing in particular that identifying $V_R$ 
and $V_{\rm CKM}$, as well as $\nu-$ and $u-$quark Yukawa couplings can
reproduce the experimental $\nu-$spectrum in a highly non-trivial way, which is 
remarkable. In sections 5 and we present the conclusions and an outlook discussing 
physical implications of these results. Finally, in the Appendix we give useful formulas concerning the eigenvalues of a (general or not) matrix.

\section{Bottom-up and Top-down parametrizations of the see-saw}

\subsection{Notation and conventions}

We will use a standard notation that can be used
for both the Standard Model (SM) and the supersymmetric (SUSY) versions of
the seesaw mechanism. The seesaw Lagrangian is given by
\bea
\label{Lseesaw}
{\cal L}\ \  \supset\ \ e_R^{c\ T} {\bf Y_e} L\cdot \bar H\  +\ 
\nu_R^{c\ T} {\bf Y} L\cdot H\ -\  {1\over 2}\nu_R^{c\ T} {\bf M} \nu_R^c\  
+\  {\rm h.c.}
\eea
where $L_i$ ($i=e,\mu,\tau$) are the left-handed lepton doublets 
(generation indices are suppressed), $(e_R^c)_i$ are the charged
lepton singlets, ${\nu_{R}}_i$ the right-handed neutrino singlets and
$H$ is the (hypercharge $=+1/2$) Higgs doublet.
${\bf Y_e}, {\bf Y}$ are the $3\times 3$ matrices of charged-leptons and
neutrino Yukawa couplings. Finally, ${\bf M}$ is a $3\times 3$ Majorana mass 
matrix for the right-handed neutrinos. Below ${\bf M}$ we can integrate out the
right-handed neutrinos, obtaining the usual
effective Lagrangian that contains a Majorana mass term for the
left-handed neutrinos:
\bea
\label{Leff}
\delta {\cal L} = - {1\over 2}\nu^T {\cal M}_\nu \nu + h.c.
\eea
where

\bea
\label{Mnu}
{\cal M}_\nu = v^2\ \kappa\  ,
\eea
with $v=\langle H^0\rangle\simeq 174\ {\rm GeV}$ and
\bea
\label{kappa}
{\bf \kappa} = {\bf Y}^T {\bf M}^{-1}{\bf Y} 
\eea
The previous equations are valid for a SUSY theory understanding all the
fields in eqs.(\ref{Lseesaw}, \ref{Leff}) as superfields, and
replacing ${\cal L}\rightarrow W$, $\delta {\cal L}\rightarrow \delta W$, 
i.e. the superpotential and the effective superpotential (with no h.c. terms).
In addition $H\rightarrow H_2$, i.e. the (hypercharge $=+1/2$) SUSY Higgs doublet
and $\langle H^0\rangle\rightarrow\langle H^0_2\rangle=v\sin\beta$, with 
$\tan \beta\equiv \langle H^0_2\rangle/\langle H^0_1\rangle$, as usual.

Working in the basis in which the charged-lepton Yukawa matrix 
(${\bf Y_e}$) and gauge 
interactions are flavour-diagonal, the neutrino mass matrix, ${\cal \kappa}$,
can be expressed as
\bea
\label{UMNS}
{\bf \kappa}= U_{\rm MNS}^*\ D_{\bf \kappa}\ U_{\rm MNS}^\dagger, 
\;\;\;\;\;\;\;D_\kappa\equiv{\rm diag}(\kappa_1, 
\kappa_2, \kappa_3),
\eea
where $\kappa_i\geq 0$ (with the convention $\kappa_1\leq 
\kappa_2\leq \kappa_3$ and thus $m_1\leq 
m_2\leq m_3$) and $U_{\rm MNS}$ is a unitary matrix
that can be written 
as\footnote{As is known, in eq.(\ref{UV}) $V$ can be multiplied from 
the left by
a diagonal unitary matrix with three independent phases. However,
these phases can be absorbed in phase redefinitions of the $e_R$ fields,
so they are no physical.}
\bea U_{\rm MNS}=V\cdot {\mathrm diag}(e^{-i\phi/2},e^{-i\phi'/2},1)\ \ ,
\label{UV}
\eea
where $\phi$ and $\phi'$ are CP violating phases (if different from
$0$ or $\pi$) and $V$ has the ordinary form of a CKM matrix
\be \label{Vdef} V=\pmatrix{c_{13}c_{12} & c_{13}s_{12} & 
s_{13}e^{-i\delta}\cr
-c_{23}s_{12}-s_{23}s_{13}c_{12}e^{i\delta} & 
c_{23}c_{12}-s_{23}s_{13}s_{12}e^{i\delta} & s_{23}c_{13}\cr
s_{23}s_{12}-c_{23}s_{13}c_{12}e^{i\delta} & 
-s_{23}c_{12}-c_{23}s_{13}s_{12}e^{i\delta} &
c_{23}c_{13}\cr}.  
\ee
Finally, note that the observable neutrino masses are given by
\bea
\label{numasses}
m_i &=& v^2 \kappa_i\;\;\;\;\;\;\;\;\;({\rm SM})
\nonumber\\
m_i &=& v^2\sin^2\beta\ \kappa_i\;\;\;\;\;\;({\rm SUSY})
\eea
Since we will be mainly interested in the $m_i/m_j=\kappa_i/\kappa_j$ 
ratios, we will work most of the time with $\kappa_i$ rather than
with $m_i$. This avoids the proliferation of annoying $v^2$, 
$v^2\sin^2\beta$ factors and permits a unified treatment of
the SM and SUSY cases [note that eq.(\ref{kappa}) is the same for both
cases]. Actually, all the results in the paper are equally valid for the SM and
the SUSY cases, except for some slight differences due to radiative effects
discussed in sect.~5.

\subsection{Basis-independent quantities}

In order to perform basis-independent analyses, it is extremely
convenient to work with basis-independent quantities. For this
matter, note that  under 
a change of basis
\bea
\label{cambiobase1}
\nu_L\rightarrow X_L \nu_L\;\;,\;\;\;\;\nu_R\rightarrow X_R \nu_R\ 
\eea
($X_{L,R}$ are arbitrary unitary matrices),
the Yukawa and mass matrices transform as
\bea
\label{cambiobase2}
{\bf Y}\rightarrow X_R^\dagger {\bf Y} X_L \;\;,\;\;\;\;  {\bf M} \rightarrow
X_R^\dagger{\bf  M} X_R^* \;\;,\;\;\;\; \kappa \rightarrow X_L^T \kappa X_L
\eea

Now the low-energy neutrino Lagrangian, eq.(\ref{Leff}), contains 
9 independent (i.e. not absorbable in field redefinitions) parameters.
They correspond to the three  mass ``eigenvalues'' $\kappa_i$
(strictly speaking they are the positive square roots of the
$\kappa\kappa^\dagger$ eigenvalues) and the six parameters
of $U_{\rm MNS}$, which is by construction a basis-independent
quantity (it is defined in a particular and well-determined basis of
the $\nu_L$ fields).

On the other hand the see-saw (high-energy) Lagrangian,
eq.(\ref{Lseesaw}), contains 18 independent parameters.
These can be defined in the following way. From eq.(\ref{cambiobase2})
is clear that one can always go to a $\nu_R$ basis where 
${\bf M}$ is diagonal, with positive entries:
\bea
\label{DiagM}
{\bf M} \rightarrow {\rm diag}(M_1, 
M_2, M_3)
\equiv D_{M} 
\eea
where we adopt the convention $M_1\leq M_2\leq M_3$. Obviously
$M_i$ are basis-independent quantities. Working in the $\nu_L$
and $\nu_R$ bases
where ${\bf Y_e}$ and ${\bf M}$, respectively, are diagonal,
the neutrino Yukawa matrix, ${\bf Y}$, can be expressed as
\bea
\label{DiagY}
{\bf Y} = V_R D_Y V_L^\dagger, \;\;\;\;\;D_Y\equiv{\rm diag}(y_1, 
y_2, y_3),
\eea
where, again, $y_i\geq 0$ and $y_1\leq y_2\leq y_3$.
The three $y_i$ parameters are obviously basis-independent quantities.
Besides $M_i$ and $y_i$, there are 12 independent high-energy
parameters contained in $V_L, V_R$. Generically, both matrices can be written
 ${\bf \Phi_1}V {\bf \Phi_2}$, where ${\bf \Phi_{1,2}}$
are diagonal unitary matrices and $V$ has the same functional form 
as (\ref{Vdef}) [replacing the $\theta_{ij}$ angles and the $\delta$
phase by new $\theta_{ij}^L$, $\delta^L$ and $\theta_{ij}^R$, $\delta^R$ 
respectively].
However, for $V_R$ the ${\bf \Phi_2}$ matrix can be absorbed into
the definition of $V_L$ [see eq.(\ref{DiagY})], so
\bea
\label{VR}
V_R = \pmatrix{e^{i\alpha_1}&&\cr&&\cr &e^{i\alpha_2}&\cr&&\cr &&1\cr} 
\pmatrix{c_{13}^R c_{12}^R  & c_{13}^R s_{12}^R  & 
s_{13}^R e^{-i\delta^R }\cr &&\cr
-c_{23}^R s_{12}^R -s_{23}^R s_{13}^R c_{12}^R e^{i\delta^R } & 
c_{23}^R c_{12}^R -s_{23}^R s_{13}^R s_{12}^R e^{i\delta^R } & s_{23}^R c_{13}^R\cr
&&\cr
s_{23}^R s_{12}^R -c_{23}^R s_{13}^R c_{12}^R e^{i\delta^R} & 
-s_{23}^R c_{12}^R -c_{23}^R s_{13}^R s_{12}^R e^{i\delta^R} &
c_{23}^R c_{13}^R} .
\eea
\\
Likewise, for $V_L$ the ${\bf \Phi_1}$ matrix can be absorbed into
phase definitions of $L=(\nu_L, e_L)^T$ and $e_R$ (keeping ${\bf Y_e}$ diagonal).
Then $V_L$ has a structure similar to $U_{\rm MNS}$ in (\ref{UV}), i.e.
$V_L = V( \theta_{12}^L, \theta_{23}^L, \theta_{13}^L, \delta^L)
\times {\rm diag}(e^{i\beta_1}, e^{i\beta_2}, 1)$.
Hence, $V_L$ and $V_R$ have 6 independent parameters each, which,
beside $M_i$ and $y_i$, complete the 18 independent parameters
of the see-saw Lagrangian\footnote{A similar discussion can be found
in ref.~\cite{Pascoli:2003uh}.}.

In summary, in the see-saw framework, the 18 (9) independent parameters 
of the high(low)-energy neutrino Lagrangian are given by the
following basis-independent quantities:

\bea
\label{list}
\begin{array}{ccc} {\rm High-Energy}& &  {\rm Low-Energy}\\
 & &\\
y_i& &\kappa_i\\
M_i& &U_{\rm MNS}\\
V_R& \longrightarrow& \\
V_L& &\\
$------------$& &$------------$\\
18\ {\rm parameters}& &9\ {\rm parameters}
\end{array}
\eea

\subsection{Bottom-up and Top-down parametrizations}

Since the number of independent parameters of the see-saw mechanism 
is larger
in the high-energy than in the effective theory, one finds often the problem
of using the available (low-energy) experimental information to constrain
the high-energy parameters. This is a bottom-up problem. 
It was shown in ref.~\cite{Casas:2001sr} that, working in the basis where 
${\bf Y_e, M}$ are diagonal and positive,
for given $D_\kappa$,
$U_{\rm MNS}$, the Yukawa matrix ${\bf Y}$ has the form
\bea
\label{param1}
{\bf Y}=D_{\sqrt{M}} R D_{\sqrt{\kappa}}U_{\rm MNS}^\dagger 
\eea
where $D_{\sqrt{M}} = \sqrt{D_M}$ (with $M_i$ arbitrary)
and $R$ is a complex orthogonal matrix (with three arbitrary complex
angles). 
Thus $D_M$ and $R$ contain the 9 additional parameters of the
high-energy theory with respect to the low-energy one.
Eq.(\ref{param1}) represents a bottom-up parametrization of the see-saw.
If desired, one can extract $y_i$, $V_L$ and $V_R$
from ${\bf Y}$ upon diagonalization.

However, for some kinds of problems it is more convenient a top-down
parametrization, i.e. a way to obtain, as directly as possible, the physical
low-energy parameters from the high-energy ones. This is precisely the sort
of problem considered here: what kind of low-energy neutrino spectrum can we
naturally expect, starting with reasonable or well-motivated choices
of the high-energy parameters\footnote{Related work on top-down parametrizations
and analysis of top-down questions can be found e.g. in ref.~\cite{Broncano:2003fq}}. 
Obviously, starting with the high-energy
parameters in (\ref{list}) one can use eqs.(\ref{DiagY}, \ref{kappa})
to write $\kappa$ in the basis where ${\bf Y_e, M}$ are diagonal
\bea
\label{seesaw}
\kappa = {\bf Y^T} D_{M^{-1}}{\bf Y} 
= V_L^* D_Y V_R^T D_{M^{-1}}V_R D_Y
V_L^\dagger \ ,
\eea
and then, upon diagonalization, determine $U_{MNS}$ and 
$\kappa_i$. Nevertheless it would be useful to find a more direct way
to extract the neutrino masses, $\kappa_i$, from the high-energy parameters.
To this end it is interesting to notice that 
$\kappa_i$ do not depend on $V_L$. In particular,
they can be obtained upon diagonalization of
\bea
\label{seesaw2}
\kappa' = D_Y V_R^T D_{M^{-1}}V_R D_Y\ , 
\eea
which is simply $\kappa$ after redefining $\nu_L$ as in 
eq.(\ref{cambiobase1}) with $X_L=V_L$. This means that
$D_{\kappa}=W_L^T \kappa' W_L$ for a certain unitary
$W_L$ matrix, or, in other words,
\bea
\label{seesaw3}
\fbox{
$D_{\kappa^2} = {\rm Eigenv}\{\kappa'\kappa'^\dagger\}$
}
\eea
Therefore, given $D_Y$ and $D_M$, the $V_R$ matrix tells the values of
$\kappa_i$. $V_L$ and $U_{\rm MNS}$ get completely decoupled from this flux 
of information: note that 1) eq.(\ref{seesaw3}) does not depend on $V_L$ and
2) the connection of $V_L$ and $U_{\rm MNS}$ is given by
\bea
\label{VLW}
\fbox{
$U_{\rm MNS} = V_L W_L$}
\eea
where $W_L$ has been defined after eq.(\ref{seesaw2}).
This means that
for {\em any} choice of $V_R$, one can always choose $V_L$ so that
the experimental $U_{\rm MNS}$ is reproduced.

Eqs.(\ref{seesaw3}, \ref{VLW}) [with $\kappa'$, $W_L$ defined in 
eq.(\ref{seesaw2}) and the lines below] represent a top-down parametrization 
of the see-saw which is useful for our purposes. 
The $V_R$ matrix, in particular,
plays here a similar role as the $R$ matrix in the bottom-up 
parametrization (\ref{param1}). They encode the flux of 
information about matrix eigenvalues
along the top-down and bottom-up directions,

\vspace{0.5cm}
\bea
\label{flujos1}
D_Y,\ D_{M}\;\;\; \stackrel{V_R}{\longrightarrow} \;\;\; D_{\kappa}
\eea
\bea
\label{flujos2}
D_\kappa,\ D_{M}\;\;\; \stackrel{R}{\longrightarrow} \;\;\; D_{Y} 
\;,
\eea
through eqs.(\ref{param1}, \ref{seesaw3}) respectively.
$U_{\rm MNS}$ gets completely decoupled from this flux of information
and can always be fitted.
[This has been just explained for the top-down parametrization.
For the bottom-up one, note
from (\ref{param1}) that $D_{Y}$ depends on $D_{\kappa}$,
$D_{M}$ and $R$, but not on $U_{\rm MNS}$.] Hence, it is not surprising
that $V_R$ and $R$ contain the same number of parameters (6 for
three families of neutrinos). The connection between them is given by
\bea
\label{YYdag}
Y Y^\dagger = D_{\sqrt{M}} R D_{\kappa} R^\dagger D_{\sqrt{M}} = V_R
D_{Y^2} V_R^\dagger  \ .
\eea
It is worth mentioning that $V_R$ has a precise
physical meaning: it measures the misalignment between ${\bf Y}$ and 
${\bf M}$. If $V_R$ is non-diagonal, there is no $\nu_R$ basis in which
${\bf Y}$ and 
${\bf M}$ can get simultaneously diagonal. The $V_R$ entries can be identified
as genuine physical inputs (and in fact they play a relevant role
in certain physical processes, as those related to leptogenesis).
On the other hand, $R$ has a more obscure physical meaning, even though
it is a useful tool for phenomenological analyses.

\section{The 2-neutrino system}

Although the case of two families of (left and right) neutrinos is obviously
non-realistic\footnote{Actually, the analysis presented in this section
is also valid for the case of three left-handed neutrinos and two right-handed neutrinos, which is the minimal version of the see-saw model capable of accommodating the low-energy observations \cite{{Frampton:2002qc}}.},
it is very useful in order to gain intuition about the form 
of the  low-energy spectrum for typical high-energy inputs.
In this case $V_R$ has the form
\bea
\label{VR2}
V_R = \left[\begin{array}{cc} e^{i\alpha}&\\&1
\end{array}\right]\ 
\left[\begin{array}{cc} c^R&s^R\\-s^R&c^R\end{array}\right]\ 
\eea
We will first obtain some simple and general relations involving
$V_R$, $D_M$, $D_Y$ and $D_\kappa$, which however
contain much information. In particular they put useful
constraints on
$V_R$ to achieve a soft normal hierarchy, $\kappa_2/\kappa_1\sim 6$,
or quasi-degeneracy, $\kappa_2/\kappa_1\sim 1$ (which for two
neutrinos is equivalent to a inverse hierarchy).
The techniques used for this general analysis will be useful for
the 3-neutrino case, to be studied in the next section.

Then we will get exact results by
solving analytically the secular equation (\ref{seesaw3}) [something
too cumbersome for three families].

\subsection{General results}

From eqs.(\ref{seesaw2}, \ref{seesaw3}) is clear that 
\bea
\label{2det}
{\rm det}\{D_\kappa\}= \kappa_1\kappa_2 = {y_1^2 y_2^2\over M_1 M_2}\ ,
\eea
which does not depend on $V_R$. On the other hand, 
the hierarchy between the physical 
masses, say $h$, can be written as
\bea
\label{2hie}
h\equiv {\kappa_2\over \kappa_1}={\kappa_2^2
\over {\rm det}\{D_\kappa\}}\ ,
\eea
so any information about $\kappa_2$ translates automatically into
$h$.
Now, using eq.(\ref{seesaw3}) we can obtain additional information 
on $\kappa_2$
from the fact that
$\kappa'\kappa'^\dagger$ is a positive hermitian matrix, which means
in particular that its largest eigenvalue is larger than any diagonal entry,
i.e.
\bea
\label{ineq1}
\kappa_2^2\geq \left(\kappa'\kappa'^\dagger\right)_{ii}
=\sum_{j=1,2}|\kappa'_{ij}|^2
=y_i^2\sum_{j=1,2}y_j^2\left|(V_R)_{ki}M_k^{-1}(V_R)_{kj}\right|^2
\ ,
\;\;\;\;\;\;\;\; i=1,2
\eea

At this point we can try an ansatz for some of the high-energy
parameters. Let us assume for the moment that the hierarchy
between $y_1$ and $y_2$
is similar to the hierarchy of Yukawa couplings observed in charged
fermions: $y_2/y_1={\cal O}(20-300)$. This means that the r.h.s. of 
eq.(\ref{ineq1}) is generically dominated by 
$\left(\kappa'\kappa'^\dagger\right)_{22}$, in particular
by the term proportional to $y_2^4$:
\bea
\label{ineq2}
\kappa_2^2\geq 
\left(\kappa'\kappa'^\dagger\right)_{22}=
{y_2^4\over M_1^2}\left|(V_R)_{12}^2
+ {M_1\over M_2}(V_R)_{22}^2\right|^2 + {\cal O}
\left( {y_1^2 y_2^2\over M_1^2} \right)
\eea
where the subdominant terms are positive.
In fact, the previous inequality is typically close to an equality:
note that from $\kappa_2^2\leq {\rm tr}(\kappa'\kappa'^\dagger)$, 
it follows that
\bea
\label{ineq3}
\kappa_2^2-\left(\kappa'\kappa'^\dagger\right)_{22}\leq
\left(\kappa'\kappa'^\dagger\right)_{11} = {\cal O}
\left( {y_1^2 y_2^2\over M_1^2} \right)\ .
\eea
Therefore, eq.(\ref{ineq2}) is an equality up to terms suppressed
by ${\cal O}({y_1^2\over y_2^2})$.\footnote{Another 
inequality for $\kappa_2^2$,
similar to eq.(\ref{ineq3}) arises from considering the
Gershgorin circle associated to 
$\left(\kappa'\kappa'^\dagger\right)_{22}$, as discussed in
Appenddix A.}

Plugging eq.(\ref{ineq2}) into eq.(\ref{2hie}), we
obtain an exact inequality for $h$,
\bea
\label{ineq4}
h={\kappa_2\over\kappa_1}\geq {y_2^2\over y_1^2}
{M_2\over M_1}
\left|(V_R)_{12}^2
+ {M_1\over M_2}(V_R)_{22}^2\right|^2 
\eea
Clearly, 
for random values of the
$V_R$ entries we expect a low-energy hierarchy 
$h={\cal O}\left({y_2^2\over y_1^2}
{M_2\over M_1}\right)$, much
stronger than that of Yukawa couplings and, of course,
than the experimental one, $h^{exp}\lsim 6$.
E.g. for $M_1\simeq M_2$ we expect $h={\cal O}( 10^{2-5})$; 
for $M_2/M_1 \sim y_2/y_1$ we expect 
$h={\cal O}(10^{3-7})$.

Consequently, either we give up the natural assumption
that
the Yukawa couplings for neutrinos present a hierarchy similar 
to the other fermions', or we accept that the $(V_R)$ 
entries are far from random. (This is already a strong
conclusion that holds for the three-generation case, as
we will see in the next section.) Let us take the second point of view
and determine the constraints on $V_R$ to achieve
degeneracy or soft hierarchy in the neutrino spectrum, 
$h\simeq 1$, $h\lsim 6$ respectively.

Let us first consider the degenerate ($h=1$) case, i.e.
$\kappa_1^2,\kappa_2^2\rightarrow y_1^2y_2^2/(M_1M_2)$.
Then, if $V_R$ has real
entries, eq.(\ref{ineq4}) requires 
 $(V_R)_{12}^4\leq {y_1^2\over y_2^2}{M_1\over M_2}
\ll 1$ [ie. $s^R\simeq 0$ in the parametrization (\ref{VR2})]. 
In addition, taking $i=1$ in (\ref{ineq1}), we get an extra inequality
for $\kappa_2$
\bea
\label{ineq11}
\kappa_2^2\geq {y_1^4\over M_1^2}\left|(V_R)_{11}^2
+ {M_1\over M_2}(V_R)_{21}^2\right|^2 .
\eea
Multiplying (\ref{ineq4}) and (\ref{ineq11}) it is straightforward to check
that the degenerate case is only obtainable when $(V_R)_{21}=0$ (i.e.
$s^R=0$) and, besides, $M_2/M_1 = y_2^2/y_1^2$.

On the other hand, if $V_R$ has complex entries [$\alpha\neq 0,\pi$
in eq.(\ref{VR2})], a cancellation inside the r.h.s. of (\ref{ineq2})
is possible (in the absence of such cancellation the previous results
essentially hold). This requires 
\bea
\label{V12c}
(V_R)_{12}^2\simeq -{M_1\over M_2}(V_R)_{22}^2\;\;,
\eea
which in turn implies
$\alpha\simeq \pm \pi/2$ in (\ref{VR2}) 
(in the next subsection
we will show that $\alpha = \pm \pi/2$ exactly\footnote{Let us mention that
$\alpha = \pm \pi/2$ does not mean maximum $CP$-violation.
On the contrary, such phase can be absorbed completely in the
definition of $D_M$ [see eg. eqs.(\ref{seesaw}, \ref{seesaw2})], 
which now contains negative, but real entries.
Hence this value of $\alpha$ does not amount to any $CP$-violation.
Nevertheless, non-trivial $CP$-violating phases can still appear
from the $V_L$ sector. These translate into $CP$-phases in
$U_{MNS}$.
}).
In addition, $M_1/M_2$ cannot be arbitrarily small. From (\ref{V12c})
we see that very small $M_1/M_2$ implies $|(V_R)_{21}|\ll 1$, 
$|(V_R)_{22}|=|(V_R)_{11}|\simeq 1$,  
which
plugged into (\ref{ineq11}) gives
$M_1/M_2\gsim y_1^2/y_2^2$, thus setting a lower bound on $M_1/M_2$.
Eq.(\ref{V12c}) tells
that, unless $M_2/M_1 = {\cal O}(1)$, the degeneracy can only
be obtained by fine-tuning $s^R$ to a very small, but {\em different
from zero}, value. (This is the case in particular for
$M_2/M_1\simeq y_2/y_1$.) For random  values of $s^R$ one
is led to a huge hierarchy between the physical masses, as expected.

Let us now say how the previous conditions are relaxed if, instead
of exact degeneracy ($h=1$), we require a soft hierarchy
($h\lsim 6$). For the real case we get a relaxed condition on 
the $M_i$ hierarchy:
$h^{-1} \lsim (M_1/M_2)(y_2^2/y_1^2)\lsim h$. The upper bound corresponds
to $s^R=0$. Otherwise a tuning of $s^R$ is required.
For the complex case, whenever a cancellation inside
eq.(\ref{ineq2}) is needed, the same condition (\ref{V12c})
is obtained, thus requiring a small and tuned value of $s^R$.
This occurs in particular for $M_2/M_1\simeq y_2/y_1$.

In summary, starting with a hierarchy of neutrino Yukawa couplings
similar to that for the charged fermions leads typically to a 
very strong hierarchy of  low-energy neutrino masses (unlike
the observed one). Nevertheless, adjusting the $V_R$ entries
it is possible to get the desired degeneracy or soft hierarchy
at low-energy. The price is a fine-tuning  between 
$y_2/y_1$, $M_2/M_1$ and $V_R$. Normally a very small,
but different from zero angle in eq.(\ref{VR2}) is required.
If nature had just two species of neutrinos we would conclude
that, unless a theoretical reason is found for this tuning,
the see-saw mechanism cannot naturally lead to the observed
low-energy neutrino spectrum if one starts with hierarchical 
neutrino Yukawa couplings similar to those of other fermions.
(This applies to the model with two right-handed neutrinos and
three left-handed neutrinos mentioned in footnote 3.)

\subsection{Some exact results}

For the 2-neutrino system, the mass eigenvalues can
be obtained from eq.(\ref{seesaw3}) in terms of the high-energy
parameters in a completely analytical way. The results are particularly
simple and illustrative for the degenerate case. Then
eq.(\ref{seesaw3}) can be written as $\kappa_{\rm d}^2{\bf 1}=
\kappa'\kappa'^\dagger$, with 
$\kappa_{\rm d}^2\equiv \kappa_1^2=\kappa_2^2=y_1^2y_2^2/(M_1M_2)$.
Consequently,
\bea
\label{eqdeg}
D_{M^{-1}}V_R D_{Y^2} V_R^\dagger D_{M^{-1}}=
\kappa_{\rm d}^2 V_R^* D_{Y^{-2}} V_R^T
\eea
Comparing the matrix entries of the two sides one concludes that
the degeneracy is only achieved when
\bea
\label{eqsexact}
\alpha =\pi/2\ ,\;\;\;\;\;\cos^2\theta^R=
\frac{M_2 y^2_2-M_1 y^2_1}{(M_1+M_2)(y^2_2-y^2_1)}\ ,
\eea
which implies in turn
\bea
\label{ineqsexact}
\frac{y^2_1}{y^2_2}\leq \frac{M_1}{M_2}\ \ .
\eea
This confirms the fact that for {\em any} choice of $y_i$, $M_i$
satisfying the inequality (\ref{ineqsexact}), there is a choice
of $V_R$ [given by eq.(\ref{eqsexact})] that produces exactly
degenerate neutrinos, $\kappa_1=\kappa_2$.

On the other hand, one can check that the degeneracy is generically
achieved thanks to a fine-tuning of the high-energy parameters.
This is illustrated for $y_2/y_1=M_2/M_1=300$ (i.e. the same hierarchy
as $u-$quarks) in Fig.~2,
which shows the mass-ratio $m_2/m_1=\kappa_{2}/\kappa_1$ 
a function of $\theta^R$ for different values of the 
$\alpha$ phase. As expected, the exact degeneracy is only possible for
$\alpha=\pi/2$ and at a very small (but different from zero) value of $\theta^R$ [see
eq.(\ref{eqsexact}) and the discussion after eq.(\ref{V12c})].
Changing $\theta^R$ and $\alpha$ from their critical values, even if very slightly,
pushes rapidly $m_2/m_1$ out from the allowed experimental region (yellow band in the
figure). For larger values of $\theta^R$, one gets 
$m_2/m_1 \rightarrow {\cal O}\left({y_2^2\over y_1^2}{M_2\over M_1}\right)$, 
in agreement with the discussion of subsect. 3.1. To this respect, notice
that in the figure only a small range of $\theta^R$ values has been
represented (for the sake of clarity).

\begin{figure}
\centerline{\vbox{
\psfig{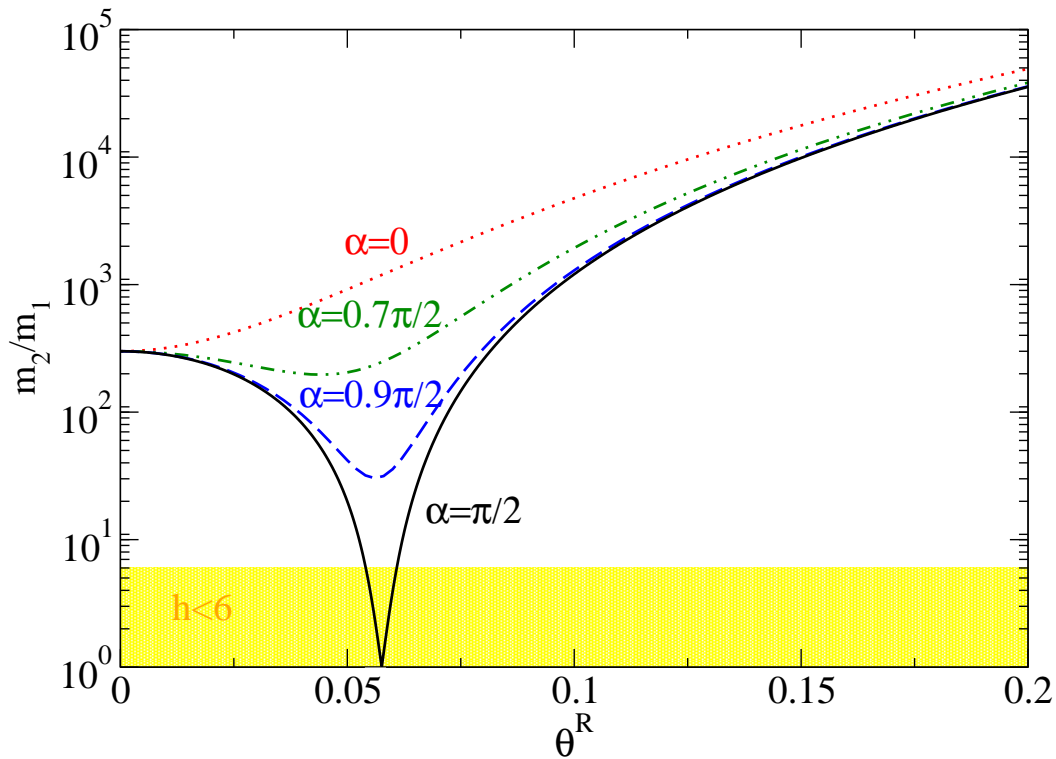}}}
\caption
{\footnotesize  
Mass-ratio $m_2/m_1$ in a 2-neutrino system vs. the $\theta^R$ angle for several
values of the $\alpha-$phase (see the notation of eq.~(\ref{VR2}))
when $y_1:y_2=M_1:M_2=1:300$.
}
\end{figure}

The conclusions are similar when $M_1\simeq M_2$, the only difference
being that the critical value of $\theta^R$ is not small.

\section{The 3-neutrino system}

Let us now examine the realistic case with three neutrino species and
a hierarchy between the two heavy ones, $h=m_3/m_2=\kappa_3/\kappa_2$,
in the experimental range: from $h\simeq 1$ (quasi-degeneracy or 
inverse hierarchy) to $h\simeq 6$ (normal but soft hierarchy).

From the results of the previous section, we can already foresee
some conclusions. First, to achieve a neutrino
spectrum where the three neutrinos are quasi-degenerate or
present a soft hierarchy will be probably as unnatural 
as for the 2-neutrino case. We will see that this indeed the case.
On the other hand, to achieve the actual
experimental constraint, namely 
soft hierarchy or quasi-degeneracy
{\em just} for the two heavy neutrinos (the latter case corresponds
to an inverse hierarchy) can be much easier.
Eg. if $V_R$ has only sizeable entries in $\{1,2\}$ box,
(i.e. $\theta_{23}=\theta_{13}=0$) 
$\kappa_1$ ($\kappa_2$) will decrease (increase) significantly,
as $\theta_{12}$ departs from zero, while $\kappa_3$ will
not change. In consequence
we expect in this case
a very large $\kappa_2/\kappa_1$ hierarchy but a softened 
$\kappa_3/\kappa_2$ one. This is consistent with
experiment and does not imply fine-tunings (only small,
but not tuned, values for certain angles).
As we will see, other possibilities can also work,
but they are not very different from the one just out-lined.

\subsection{General results}

Let us recall that the neutrino masses, $\kappa_i$, depend on the
high-energy parameters, $y_i, M_i, V_R$ through eq.(\ref{seesaw3}).
As for the 2-neutrino case, the determinant 
\bea
\label{3det}
{\rm det}\{D_\kappa\}= \kappa_1\kappa_2\kappa_3 = 
{y_1^2 y_2^2 y_3^2\over M_1 M_2 M_3}\ ,
\eea
does not depend on $V_R$. The hierarchy between the 
two heavy neutrino masses can be written as
\bea
\label{3hie}
h= {m_3\over m_2}={\kappa_3\over \kappa_2}={\kappa_3^2 \kappa_1
\over {\rm det}\{D_\kappa\}}\ ,
\eea
Now, in order to get information about $h$ we need information
on $\kappa_3, \kappa_1$.

Using the fact [eq.(\ref{seesaw3})] that $\kappa_i^2$ are the eigenvalues of 
$\kappa'\kappa'^\dagger$, which is a positive hermitian matrix,
we can write
\bea
\label{3ineq1}
\kappa_3^2\geq \left(\kappa'\kappa'^\dagger\right)_{ii}
=\sum_{j=1,2,3}|\kappa'_{ij}|^2=
y_i^2\sum_{j=1,2,3}y_j^2|(V_R)_{ki}M_k^{-1}(V_R)_{kj}|^2
\ ,
\;\;\;\;\;\;\;\; i=1,2,3
\eea
At this point we can try again an ansatz for the spectrum of high-energy
parameters. So let us assume for the moment that the hierarchy
between the $y_i$ 
is similar to the hierarchy of Yukawa couplings observed in charged
fermions: $y_3/y_2,\ y_2/y_1={\cal O}(20-300)$. Then eq.(\ref{3ineq1}) 
is generically dominated by 
$\left(\kappa'\kappa'^\dagger\right)_{33}$, in particular
by the term proportional to $y_3^4$, which corresponds to $i=j=3$:
\bea
\label{3ineq2}
\kappa_3^2\geq 
\left(\kappa'\kappa'^\dagger\right)_{33}=
{y_3^4\over M_1^2}\left|(V_R)_{13}^2
+ {M_1\over M_2}(V_R)_{23}^2+ {M_1\over M_3}(V_R)_{33}^2\right|^2 
 + {\cal O}
\left( {y_3^2 y_2^2\over M_1^2} \right)
\eea
where the subdominant terms are positive.
As for two neutrinos, the previous inequality is typically close to 
an equality: from 
$\kappa_3^2\leq {\rm tr}(\kappa'\kappa'^\dagger)$, it follows that
\bea
\label{3ineq3}
\kappa_3^2-\left(\kappa'\kappa'^\dagger\right)_{33}\leq
\left(\kappa'\kappa'^\dagger\right)_{11}
+\left(\kappa'\kappa'^\dagger\right)_{22} = {\cal O}
\left( {y_3^2 y_2^2\over M_1^2} \right)\ ,
\eea
so (\ref{3ineq2}) holds as an equality up to 
${y_2^2\over y_3^2}$--suppressed 
terms\footnote{An inequality similar to (\ref{3ineq3})
arises from the  Gershgorin theorem, as discussed in Appendix A.}.

On the other hand we can obtain information on $\kappa_1$ by considering
$\kappa'^{-1}(\kappa'^{-1})^\dagger$, which is a positive hermitian matrix
with $\kappa_i^{-2}$ eigenvalues. The largest eigenvalue, $\kappa_1^{-2}$,
satisfies
\bea
\label{3ineq1inv}
\kappa_1^{-2}\geq \left(\kappa'^{-1}(\kappa'^{-1})^\dagger\right)_{ii}
=\sum_{j=1,2,3}|\kappa'^{-1}_{ij}|^2=
y_i^{-2}\sum_{j=1,2,3}y_j^{-2}|(V_R)_{ki}M_k(V_R)_{kj}|^2
,
\;\;\; i=1,2,3
\eea
This equation is typically dominated by $\left(\kappa'^{-1}
(\kappa'^{-1})^\dagger\right)_{11}$, in particular by the $i=1, j=1$
term,
\bea
\label{3ineq1inv2}
\kappa_1^{2}\leq 
{y_1^{4}\over M_3^2}\left|(V_R)_{31}^2 +{M_2\over M_3}(V_R)_{21}^2 
+{M_1\over M_3}(V_R)_{11}^2
\right|^{-2} 
- {\cal O}
\left( {y_1^6 \over y_2^2 M_3^2} \right)
\eea
where the subdominant terms are negative (so ignoring them still
represents an exact inequality). Again, 
this inequality is typically close to an equality:
from $\kappa_1^{-2}\leq {\rm tr}\left[\kappa'^{-1}
(\kappa'^{-1})^\dagger\right]$ it follows 
that\footnote{Once more, an inequality similar to (\ref{trk11})
arises from the  Gershgorin theorem, see Appendix A.}
\bea
\label{trk11}
\kappa_1^2 \geq  \left[\sum_{i,j=1,2,3}|\kappa'^{-1}_{ij}|^2\right]^{-1}
\ ,
\eea
which is dominated by $i=j=1$:
\bea
\label{trk12}
\kappa_1^{2}\geq 
{y_1^{4}\over M_3^2}\left|(V_R)_{31}^2 +{M_2\over M_3}(V_R)_{21}^2 
+{M_1\over M_3}(V_R)_{11}^2
\right|^{-2} 
- {\cal O}
\left( {y_1^6 \over y_2^2 M_3^2} \right)
\eea
Note that in eq.(\ref{trk12}) the subdominant terms are negative
(so ignoring them here represents an {\em approximate} inequality).
In any case, comparing (\ref{3ineq1inv2}) and (\ref{trk12}),
we see that eq.(\ref{3ineq1inv2}) holds
as an equality up to 
${y_1^2\over y_2^2}$--suppressed 
terms.

Similarly to the 2-neutrino case, 
plugging eqs.(\ref{3ineq2}, \ref{trk12})
into eq.(\ref{3hie}) we get an inequality\footnote{Plugging
eq.(\ref{trk11}) instead of eq.(\ref{trk12}) into eq.(\ref{3hie})
we obtain an {\em exact} inequality for $h$, though slightly more involved than 
(\ref{haprox}). On the other hand, a simpler approximate inequality
is obtained from (\ref{haprox}) by noting that the absolute value in the
denominator is $\leq 1$.} for $h$,
\bea
\label{haprox}
h= {m_3\over m_2}={\kappa_3\over \kappa_2}\ \gsim \ 
{y_3^2\over y_2^2}{M_2\over M_1}
{\left|(V_R)_{13}^2
+ {M_1\over M_2}(V_R)_{23}^2+ {M_1\over M_3}(V_R)_{33}^2\right|^2 
\over
\left|(V_R)_{31}^2 +{M_2\over M_3}(V_R)_{21}^2 
+{M_1\over M_3}(V_R)_{11}^2
\right| }\ .
\eea
From this expression it is clear that
for random values of the
$V_R$ entries we expect a low-energy hierarchy
much
stronger than that of Yukawa couplings.

Only for $y_3/y_2$, $M_2/M_1$ $={\cal O}(1)$ can the experimental
value $h^{exp}\lsim 6$ be naturally obtained.
For a $y_i/y_j$ hierarchy similar to quarks and charged leptons,
we expect  we expect $h={\cal O}( 10^{2-5})$ if $M_1\simeq M_2\simeq M_3$,
and $h={\cal O}(10^{3-7})$ if $M_i/M_j \sim y_i/y_j$ (which is 
probably a more attractive possibility), in any case way too large.

So we arrive to a similar conclusion as for two neutrinos:
either we give up the natural assumption
that the neutrino Yukawa couplings present a hierarchy 
similar to other fermions, or we accept that the $V_R$ 
entries are far from random.
However, in this case ``far from
random'' does not necessarily mean ``fine-tuned'', as will be
shown in subsect. 4.3. 

We will devote subsects. 4.2 and 4.3
to determine the pattern of
$V_R$ required to achieve the desired 
soft hierarchy (or quasi-degeneracy) for the three
neutrinos or just for the two heavy ones respectively.
Let us advance that since the absolute value
in the denominator of (\ref{haprox}) is $\leq 1$, 
then $\left|(V_R)_{13}^2
+ {M_1\over M_2}(V_R)_{23}^2+ {M_1\over M_3}(V_R)_{33}^2\right|^2\ll 1$
must be fulfilled in all cases.

\vspace{0.2cm}
\noindent
{\underline {\bf Connection with models of anarchic neutrinos}}

We would like to make a very short digression about the use of the previous
approach to analyze scenarios of anarchic neutrinos \cite{Hall:1999sn}. The
basis-independent top-down formulation of the see-saw mechanism that we are using
may be convenient to make statistical considerations about the
high-energy parameters that define the theory, as is done in
models of anarchic neutrinos. In particular, in the absence
of additional assumptions, it makes sense
to scan $y_i$ and the 6 parameters defining $V_R$ instead of
the ${\bf Y}$ matrix, which contains 18 parameters (3 of them redundant and 6
not related to the neutrino masses).

Then, from (\ref{haprox}) we notice that for average values of the $V_R$ entries,
in particular for $|(V_R)_{13}|^2 \sim |(V_R)_{31}|^2\sim 1/3$,
we get a hierarchy $h\sim {1\over 3}{y_3^2\over y_2^2}{M_2\over M_1}$.
Therefore the expectable pattern of neutrino masses depends crucially
on the range in which the $y_i, M_i$ parameters are allowed to vary.
E.g. if one uses $y_i\in [1/a, a]y_0$,
$M_i\in [1/a, a]M_0$, with $a>1$, one expects $h\sim a^3/3$.

\subsection{Degeneracy or soft hierarchy for the three neutrinos}

Let us first consider the case of completely degenerate 
low-energy neutrinos. From eq.(\ref{3det}) this means

\bea
\label{kdeg}
\kappa_1^2=\kappa_2^2 =\kappa_3^2\equiv\kappa_{\rm deg}^2=
\left({y_1^2y_2^2y_3^2\over M_1M_2M_3}\right)^{2/3}
\eea
Now we will use the inequalities (\ref{3ineq1}, \ref{3ineq1inv})
for $i=j=1,3$. This produces four inequalities, which are given
by eqs.(\ref{3ineq2}, \ref{3ineq1inv2}) and

\bea
\label{inequal3}
\kappa_3^2\geq 
{y_1^{4}\over M_3^2}\left|(V_R)_{31}^2 
+ {M_3\over M_2}(V_R)_{21}^2 + {M_3\over M_1}(V_R)_{11}^2 \right|^2
\eea
\bea
\label{inequal4}
\kappa_1^2\leq {y_3^{4}\over M_1^2}
\left|(V_R)_{13}^2
+ {M_2\over M_1}(V_R)_{23}^2+ {M_3\over M_1}(V_R)_{33}^2\right|^{-2}
\eea
Again we assume a strong  hierarchy
among the $y_i$, say  
similar to the hierarchy of Yukawa couplings observed in charged
fermions: $y_3/y_2,\ y_2/y_1={\cal O}(20-300)$. We do not assume
a priori any particular hierarchy between the three $M_i$, except the
conventional ordering $M_1\leq M_2\leq M_3$.

Let us suppose for the moment that there are no delicate cancellations
among the terms in the right-hand sides of eqs.(\ref{3ineq2}, \ref{3ineq1inv2},
\ref{inequal3}, \ref{inequal4}). This means that the absolute value of each
term inside the straight brackets is $\lsim$ the absolute value of the sum
of them (note that ``$\lsim$'' becomes ``$\leq$'' for real $V_R$).
Then, 
since ${y_3^{2}/M_1}\gg \kappa_{\rm deg}$, ${y_1^{2}/M_3}\ll \kappa_{\rm deg}$,
it is clear from eqs.(\ref{3ineq2}) and (\ref{3ineq1inv2})
that $|(V_R)_{13}|^2\ll 1$ and $|(V_R)_{31}|^2\ll 1$ respectively.
Besides, the unitarity of $V_R$ implies that either {\em a)} $|(V_R)_{23}|^2,\
|(V_R)_{32}|^2 \ll 1$
or {\em b)} $|(V_R)_{21}|^2,\ |(V_R)_{12}|^2\ll 1$, i.e. $V_R$ is approximately
box-diagonal. Furthermore, looking at the $(V_R)_{ij}^2$-term with 
smaller factor
in eqs.(\ref{3ineq2}) and (\ref{3ineq1inv2}) we obtain
\bea
\label{inequal5}
{y_3^{2}\over M_3}\lsim \kappa_{\rm deg}\ ,
\;\;\;\;\;\;\;\;{y_1^{2}\over M_1}\gsim \kappa_{\rm deg}
\eea
respectively. This implies $M_3/M_1\ \gsim y_3^2/y_1^2$. (This works 
similar to the case of two neutrinos, see eq.(\ref{ineqsexact}).)
Suppose $V_R$ falls in the possibility {\em a)} above, which means 
$|(V_R)_{33}|^2={\cal O}(1)$. Then eq.(\ref{inequal4})
implies $y_3^{2}/M_3\gsim \kappa_{\rm deg}$ which, together
with the first equation in (\ref{inequal5}), requires 
\bea
\label{autov1}
{y_3^{2}\over M_3}\simeq \kappa_{\rm deg}
\eea
This corresponds to the fact that $V_R$ is essentially diagonal, except
in the 1-2 box. Eqs.(\ref{autov1}, \ref{kdeg}) imply 
${y_1^{2}y_2^{2}\over M_1M_2}\simeq\kappa_{\rm deg}^2$. 
Due to the large $y-$hierarchy,
this means ${y_1^{2}\over M_2}\ll \kappa_{\rm deg}\ll {y_2^{2}\over M_1}$.
Applying this to the second term in the r.h.s. of eq.(\ref{3ineq1inv2}),
we conclude $|(V_R)_{21}|^2\ll 1$ (and $|(V_R)_{11}|^2\simeq 1$, 
$|(V_R)_{12}|^2\ll 1$ 
by unitarity). So $V_R$ is essentially ${\bf 1}$. Actually, from the
third term of (\ref{inequal3}) we obtain
${y_1^{2}\over M_1}\lsim \kappa_{\rm deg}$, which together with
 eq.(\ref{inequal5}), implies 
${y_2^{2}\over M_2}\simeq \kappa_{\rm deg}$. Had we started with
the possibility {\em b)} above, we would have obtained the same conclusion.
In summary, if there are no precise cancellations in the r.h.s.
of eqs.(\ref{3ineq2}, \ref{3ineq1inv2},
\ref{inequal3}, \ref{inequal4}), the only choice of high-energy 
parameters giving completely degenerate neutrinos is
\bea
\label{sol-no-can}
{y_1^{2}\over M_1}\simeq {y_2^{2}\over M_2}\simeq {y_3^{2}\over M_3}
\simeq\kappa_{\rm deg}^2\ ,
\;\;\;\;\;\;V_R\simeq {\bf 1}
\eea
This is similar to the 2-neutrino case.

If the $y_i, M_i, V_R$ parameters are not in the relation 
(\ref{sol-no-can}), we are forced to admit non-trivial cancellations 
between the various terms in the right-hand-sides of 
eqs.(\ref{3ineq2}, \ref{3ineq1inv2},
\ref{inequal3}, \ref{inequal4}). In particular, if such cancellation
exists in the r.h.s. of eq.(\ref{3ineq2}) {\em and} eq.(\ref{3ineq1inv2}),
the constraints (\ref{inequal5}) [and the subsequent
$M_3/M_1\ \gsim y_3^2/y_1^2$ inequality] do not apply. Actually, for a wide
range of $y_i, M_i$ parameters, the entries of $V_R$ can be arranged
so that the two cancellations take place and
$\kappa_1=\kappa_2 =\kappa_3=\kappa_{\rm deg}$
(see below for more details).
However,
this amounts to a very accurate (and thus unplausible) fine-tuning.
This result cannot be easily appreciated if one just uses the bottom-up
see-saw parametrization, eq.(\ref{param1}), since this automatically
gives sets of working $y_i$ parameters for arbitrary $M_i, R$.
An intermediate situation occurs when the cancellation takes place
``just'' in one of the right-hand-sides of eqs.(\ref{3ineq2}, \ref{3ineq1inv2},
\ref{inequal3}, \ref{inequal4}). Eg. suppose that the cancellation
just occurs in the r.h.s. of eq.(\ref{3ineq1inv2}). Then, from 
eqs.(\ref{3ineq2}, \ref{inequal4}) we easily conclude that
${y_3^{2}\over M_2}|(V_R)_{23}|^2\lsim \kappa_{\rm deg}
\lsim {y_3^{2}\over M_3}(1-|(V_R)_{23}|^2)^{-1}$, which implies
that either ${y_3^{2}\over M_3}$ or ${y_3^{2}\over M_2}$ must be close
to $\kappa_{\rm deg}$. 

In any case, we have seen that unless the high-energy parameters 
satisfy (\ref{sol-no-can}), fine cancellations are required
in order
to obtain degenerate neutrinos. 
Then, in the absence of an explanation
for such cancellations, we conclude that degenerate neutrinos 
are not natural within the see-saw framework if the neutrino
Yukawa couplings present a hierarchy similar to other 
fermions\footnote{See ref.\cite{Ramond} for
the discussion of a particular theoretical model}. Let us also note that
sometimes is stated that (see-saw) degenerate neutrinos 
naturally require degenerate right-handed Majorana masses, $M_i$,
as well. Now we see that this
is only true if the Yukawa couplings are degenerate as well, according
to eq.(\ref{sol-no-can}). Otherwise a fine-tuning for the $V_R$
entries is needed, exactly as for other choices of $M_i$.

Let us now be more precise about what conditions must fulfill the $y_i, M_i$
parameters in order to exist a choice of $V_R$ that implements
degenerate neutrinos. First of all, notice that if 
${y_2^{2}\over M_2}=\kappa_{\rm deg}$, the problem reduces
to a 2-neutrino one, in this case the 1 and 3 neutrinos.
[This occurs in particular when both the $y_i-$ and the
$M_i-$hierarchies are regular, i.e. $y_3/y_2=y_2/y_1$,
$M_3/M_2=M_2/M_1$.]
Then, from the results of the previous section, we know that,
provided $M_3/M_1 \leq y_3^2/y_1^2$, there will be a non-trivial
solution. The corresponding $V_R$ matrix is non-trivial
in the 1--3 box.
Since the 
$y_3^2/y_1^2$ ratio is normally very large, the fine-tuning
in the values of the $V_R$ entries must be extremely precise.
More generally, 
we can obtain necessary conditions for
$y_i, M_i$ in order to accommodate degenerate neutrinos
as follows.
Using the  bottom-up
see-saw parametrization (\ref{param1}), if neutrinos are degenerate 
we can write
\bea
Y Y^\dagger = \kappa_{\rm deg} \sqrt{D_M} R R^\dagger \sqrt{D_M}
\eea
where $\kappa_{\rm deg}$ is given by (\ref{kdeg}).
Since $Y Y^\dagger$ is a positive hermitian matrix,
its largest eigenvalue, $y_3^2$, must be
larger than the diagonal entries, i.e.
\bea
y^2_3\geq \kappa_{\rm deg}\{ M_i (R R^\dagger)_{ii}\},\;\;\;\;i=1,2,3
\label{cotay2}
\eea 
Taking into account $(R R^\dagger)_{ii}\geq 1$ (this can be readily
checked using eg. the parametrization of $R$ given in ref.~\cite{Casas:2001sr})
we finally obtain
\bea
y^2_3 \geq \kappa_{\rm deg} M_3
\label{cond1}
\eea
A similar argument applied to the $(Y Y^\dagger)^{-1}$ matrix leads to
\bea
y^2_1 \leq \kappa_{\rm deg} M_1
\label{cond2}
\eea
Note that eqs.(\ref{cond1}, \ref{cond2}) imply $M_3/M_1\leq y_3^2/y_1^2$.
Let us stress that these are necessary
but not sufficient conditions to guarantee the existence of a 
$V_R$ matrix producing degenerate neutrinos. Nevertheless the 
numerical analysis shows that in most cases satisfying the above
conditions such $V_R$ matrix can be found. Note that conditions
(\ref{cond1}, \ref{cond2}) are only compatible with 
the constraints (\ref{inequal5}) [obtained under the assumption
of no fine-tunings in $V_R$] when eq.(\ref{sol-no-can}) is fulfilled,
in agreement with the previous discussion.

In summary, if neutrino Yukawa couplings present a hierarchy similar
to other fermions, a spectrum of completely degenerate (or quasi-degenerate)
neutrinos is possible but
quite unnatural. For random $V_R$ the hierarchy of neutrino masses
is actually much stronger than that of Yukawa couplings, in absolute
conflict with experimental data. For $V_R={\bf 1}$ a degenerate spectrum
if the Yukawa couplings, $y_i$, and the right-handed masses, $M_i$ are
in the precise proportion (\ref{sol-no-can}). For arbitrary $y_i$, $M_i$ 
satisfying (\ref{cond1}, \ref{cond2}) it is in general possible to 
find a particular $V_R$ giving degenerate neutrinos, but this amounts
to a strong fine-tuning. 

Finally, let us remark that these conclusions
still hold (although somewhat softened) if instead degenerate neutrinos
one demands hierarchical neutrinos with a soft hierarchy between {\em the
three} families, e.g. $\kappa_3/\kappa_2\lsim 6$
(this is obliged by experimental data) {\em and} $\kappa_2/\kappa_1\lsim 6$
(this is just an hypothesis).

These results strongly suggest to consider soft hierarchy or quasi-degeneracy
{\em just} for the two heavy neutrinos, which we study next.

\subsection{Degeneracy or soft hierarchy for ${\bf m_3/m_2}$}

We will focus now on the possibility of fulfilling 
$h={m_3\over m_2}={\kappa_3\over \kappa_2}\lsim 6$
(i.e. the only experimental constraint on the ratio of neutrino masses), 
starting with hierarchical Yukawa couplings. Again we will assume
for the moment that the hierarchy
between the $y_i$ 
is similar to the hierarchy of Yukawa couplings observed in charged
fermions: $y_3/y_2,\ y_2/y_1={\cal O}(20-300)$.

For convenience for the discussion we repeat here the
previous bound (\ref{haprox}) on the value of $h$,
\bea
\label{haproxp}
h\gsim
{y_3^2\over y_2^2}{M_2\over M_1}
{\left|(V_R)_{13}^2
+ {M_1\over M_2}(V_R)_{23}^2+ {M_1\over M_3}(V_R)_{33}^2\right|^2 
\over
\left|(V_R)_{31}^2 +{M_2\over M_3}(V_R)_{21}^2 
+{M_1\over M_3}(V_R)_{11}^2
\right| }\ ,
\eea
As discussed in subsect.4.1, this equation tells us that for random
values of the $V_R$ entries we expect 
$h\sim{y_3^2\over y_2^2}{M_2\over M_1}\gg 6$. Therefore we need
to imagine ways to get $h$ much smaller than the ``random'' result,
preferably without fine-tunings. Obviously this is much easier 
to achieve if the combination of $V_R$ elements in the denominator 
of (\ref{haproxp})
is as large as possible. From (\ref{3ineq1inv2}) this corresponds 
to $\kappa_1$ as small as possible. Therefore generically it is far 
more natural to get the experimental result $h\lsim 6$ if the 
lightest neutrino presents a much stronger hierarchy than the two
heavy ones, which is an interesting conclusion\footnote{An exception
to this rule occurs when the $y_i, M_i$ values are in the proportion
(\ref{sol-no-can}). Then $V_R\sim {\bf 1}$ leads naturally to
degenerate or soft-hierarchical neutrinos.}.

However, a denominator as large as possible is not enough to render
$h\lsim 6$: the expression in straight brackets in the denominator
is $\leq 1$, so a small numerator, $\left|(V_R)_{13}^2
+ {M_1\over M_2}(V_R)_{23}^2+ {M_1\over M_3}(V_R)_{33}^2\right|^2\ll 1$,
is always obliged.
If $M_1=M_2=M_3$ (i.e. degenerate right-handed
neutrinos) this can only be accomplished by a cancellation between
the various terms in the numerator. On the contrary, if
$M_1\ll M_2\ll M_3$ this could be achieved without cancellations.
We examine next the two cases separately.

\vspace{0.3cm}
\noindent
\underline{${\bf M_1\ll M_2\ll M_3}$}

\vspace{0.2cm}

\noindent

{\em If} we do {\em not} allow fine cancellations in the numerator of
eq.(\ref{haproxp}), this gets minimal when is dominated by
the $(V_R)_{33}^2$ term. This requires $|(V_R)_{13}|^2$ and
$|(V_R)_{23}|^2$ $\ll 1$;\footnote{If the hierarchy of $M_i$ is
very strong, the dominance of the $(V_R)_{33}^2$ term may be
non-compulsory.
More precisely, if $M_1/M_2\lsim y_3^2/y_2^2$, then the condition
$|(V_R)_{23}|^2\ll 1$ can be relaxed ($|(V_R)_{13}|^2\ll 1$ cannot).}
more precisely  $|(V_R)_{13}|^2< M_1/M_3$
and $|(V_R)_{23}|^2< M_2/M_3$. Then $(V_R)_{31}^2$ in
the denominator is also very small (by unitarity of $V_R$) and 
eq.(\ref{haproxp}) can normally be approximated as
\bea
\label{haproxp2}
h \gsim
{y_3^2\over y_2^2}{M_1\over M_3}
{1 \over 
|(V_R)_{21}^2+{M_1\over M_2}(V_R)_{11}^2|}  \ .
\eea
In particular,
\bea
\label{haproxp3}
h \gsim 
{y_3^2\over y_2^2}{M_1\over M_3}
{1 \over 
|(V_R)_{21}|^2}  \;\;\; \;\;\;\;\;\;{\rm if\;\;\; }
|(V_R)_{21}|^2\gsim (M_1/M_2) .
\eea
\bea
\label{haproxp4}
h \gsim 
{y_3^2\over y_2^2}{M_2\over M_3}
\;\;\;\;\;\;\;\;\;\;\;\;\;\;\;\;\;\;\;\;\;\;\;{\rm otherwise}\ \ .
\eea
E.g. if ${M_i/ M_j}= {y_i/ y_j}$ (which we find a reasonable assumption) 
in a regular hierarchy, i.e. ${y_1\over y_2}={y_2\over y_3}$, then 
the case (\ref{haproxp3}) becomes
$h\lsim |(V_R)_{21}|^{-2}$. As a matter of fact, 
taking 
$|(V_R)_{21}|=1$ and $(V_R)_{11}$, $(V_R)_{13}=0$, leads exactly
to $\kappa_3 = \kappa_2$ and thus
inverse hierarchy. This can be easily checked using the exact results
of subsect.3.2, since in this limit the problem involves only two neutrinos. 
More generally, for sizeable $|(V_R)_{21}|^2$ we get a soft hierarchy
for the two heavy neutrinos. E.g. for $|(V_R)_{21}|^2\gsim 1/6$ we get
$h\lsim 6$, in agreement with experiment.
Notice that there are no delicate cancellations (and thus no fine-tuning)
involved in this instance: changes in the $V_R$ entries 
amount to changes in $h$ in a similar proportion.
On the other hand, for very small $|(V_R)_{21}|$ 
(and thus very small $|(V_R)_{12}|$ by unitarity)
eq.(\ref{haproxp2})
becomes $h\gsim {y_3^2\over y_2^2}{M_2\over M_3}= {y_3\over y_2}$, which is 
too large.

Let us stress that the above possibility of getting an experimentally viable
$h$ with no fine-tunings requires very small 
$|(V_R)_{13}|$, $|(V_R)_{23}|$, and sizeable $|(V_R)_{21}|$.\footnote{
An  intuitive way to understand the pattern obtained for $V_R$
is to realize that
it simply corresponds to a ``random" $2\times 2$ box for the two lighter
neutrinos and the rest close to the identity matrix. Then $\kappa_1$
 and $\kappa_2$ split enormously, as shown in sect.3, and thus
$\kappa_2$ approaches $\kappa_3$ (which changes little), while
$\kappa_1$ gets extremely small.}
This coincides exactly with the structure of the CKM matrix, which
we find very suggestive. Actually, the coincidence is even stronger
since the previous discussion suggests
$|(V_R)_{13}|^2\ll |(V_R)_{23}|\ll |(V_R)_{21}|=$ sizeable, as for
CKM. We will turn to a more careful exam of this CKM-like form
for $V_R$ in sect.5.

Another (less attractive) possibility to get a small numerator in 
eq.(\ref{haproxp}) is to allow for cancellations between the various terms
inside the straight brackets. This requires $|(V_R)_{13}|^2\gsim M_1/M_3$
and/or $|(V_R)_{23}|^2\gsim M_2/M_3$. Still, this possibility
requires very small $|(V_R)_{13}|$. The largest possible
value for  $|(V_R)_{13}|$ occurs when it cancels against the 
$(V_R)_{23}^2$ term, so 
\bea
\label{V13bound}
|(V_R)_{13}|^2\lsim  {M_2\over M_3}\ .
\eea
These results are illustrated in Fig.~3, where we
show the density of allowed points in the $|(V_R)_{13}|-|(V_R)_{23}|$
plane for fixed values of $|(V_R)_{12}|$ [this determines the $V_R$ matrix
up to phases, according to eq.(\ref{VR})] and
$y_1:y_2:y_3$ $=1:300:9\times 10^4$,
$M_1:M_2:M_3$ $=1:300:9\times 10^4$.  In each point, we have evaluated
$h$ for 1000 random values of the phases
in $V_R$, and counted the number of points that are compatible
with the observed hierarchy, $h\lsim 6$. White areas are excluded, while
colored areas are allowed, corresponding the redder (darker in black and white
printer) areas
to the regions with higher density of allowed points. The
reddest areas precisely correspond to the choices of $V_R$ that
reproduce naturally (with no cancellations) the observed mass hierarchy. 
As discussed just before eq.(\ref{haproxp2}), this occurs for 
$|(V_R)_{13}|^2< M_1/M_3$
and $|(V_R)_{23}|^2< M_2/M_3$,
thus the size and shape of the reddest ``rectangle''.
The light blue (light grey in black and white) areas correspond to the choices of $V_R$ that
can reproduce the observations with a certain
amount of tuning. As argued above, for small
$|(V_R)_{12}|$ it is not possible to reproduce $h^{exp}$, 
unless a fine-tuning in the numerator of (\ref{haproxp})
takes place, thus the tiny light allowed areas 
for $|(V_R)_{12}|^2\lsim 1/6$, in agreement with 
the previous discussion. The bound (\ref{V13bound})
is also clearly visible.

\begin{figure}
\hspace{-2.5cm}
\begin{tabular}{c}
\psfig{figure=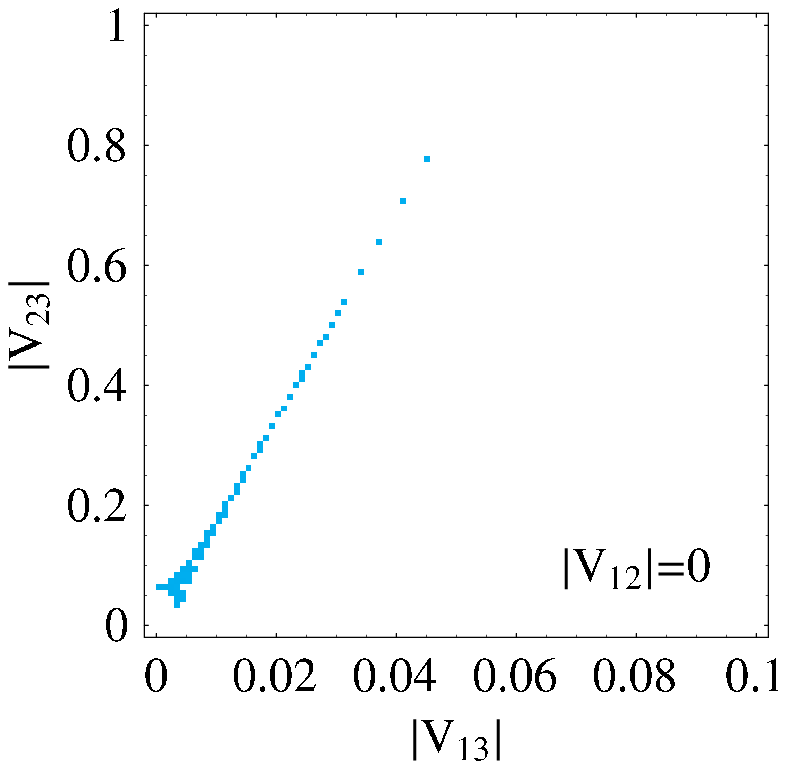,width=90mm}  \hspace{-2cm}
\psfig{figure=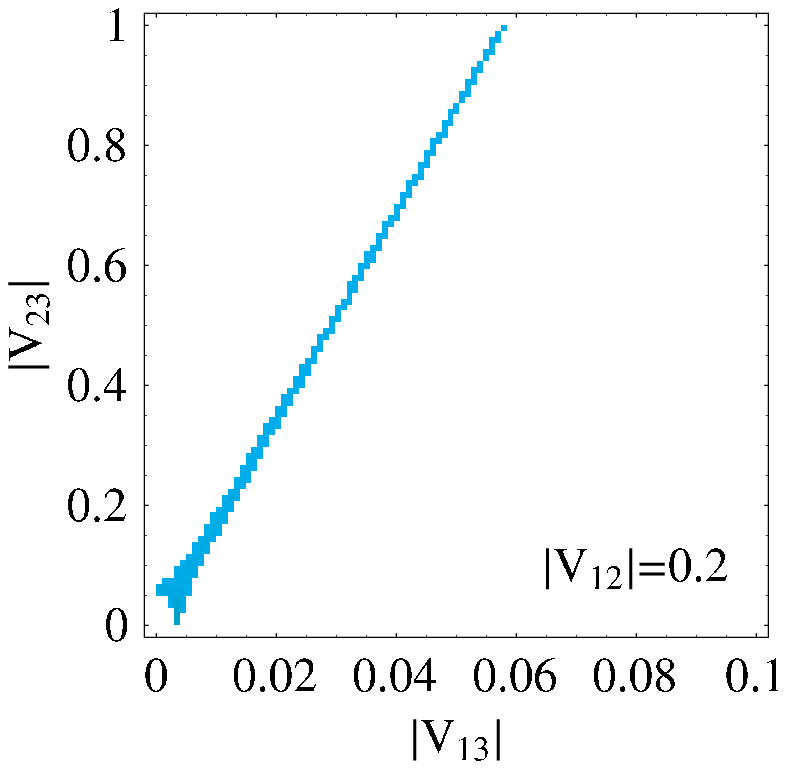,width=90mm} \\
\psfig{figure=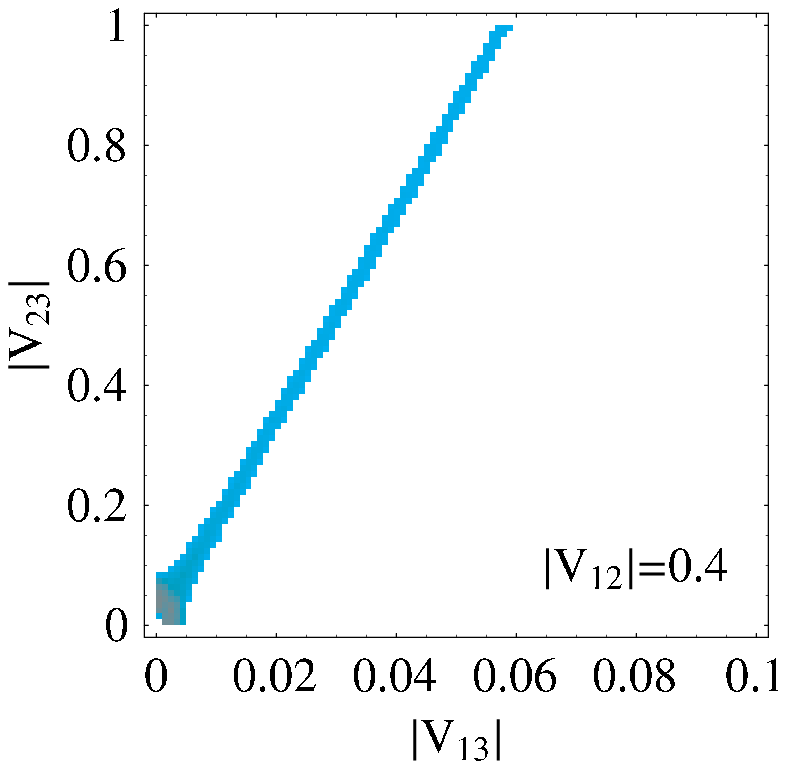,width=90mm}   \hspace{-2cm}
\psfig{figure=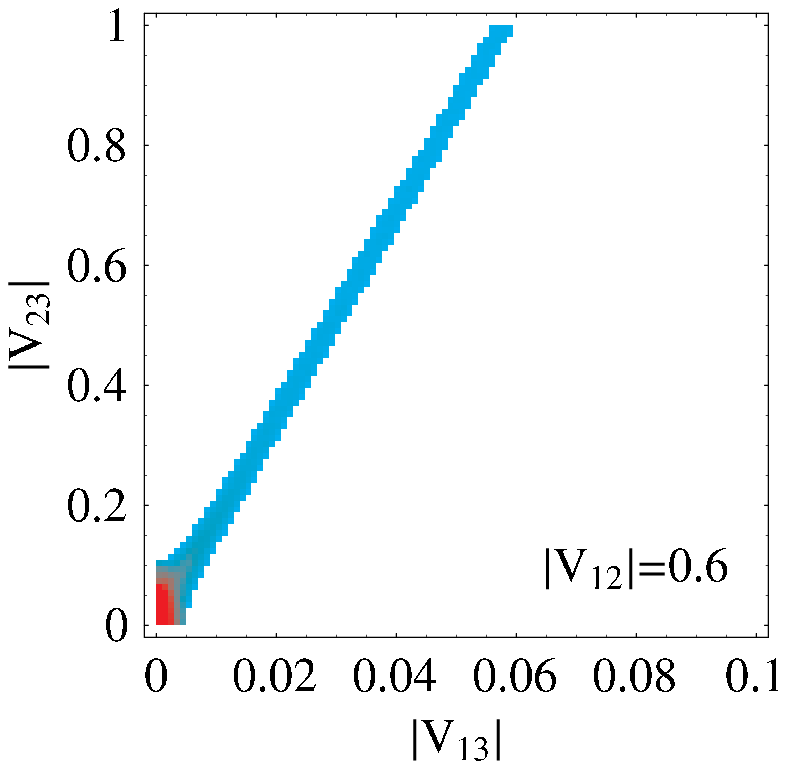,width=90mm} \\
\psfig{figure=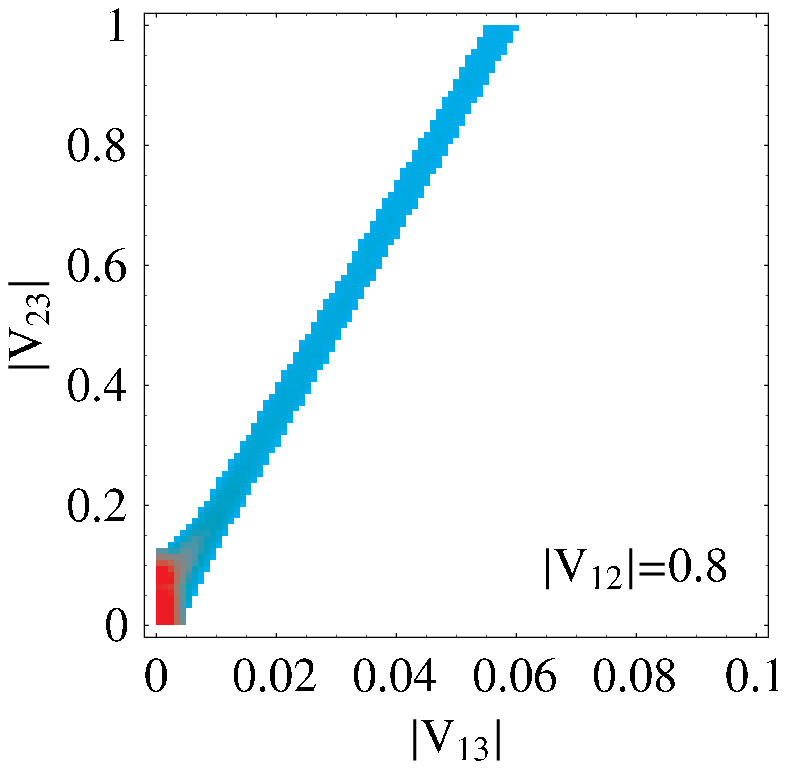,width=90mm}  \hspace{-2cm}
\psfig{figure=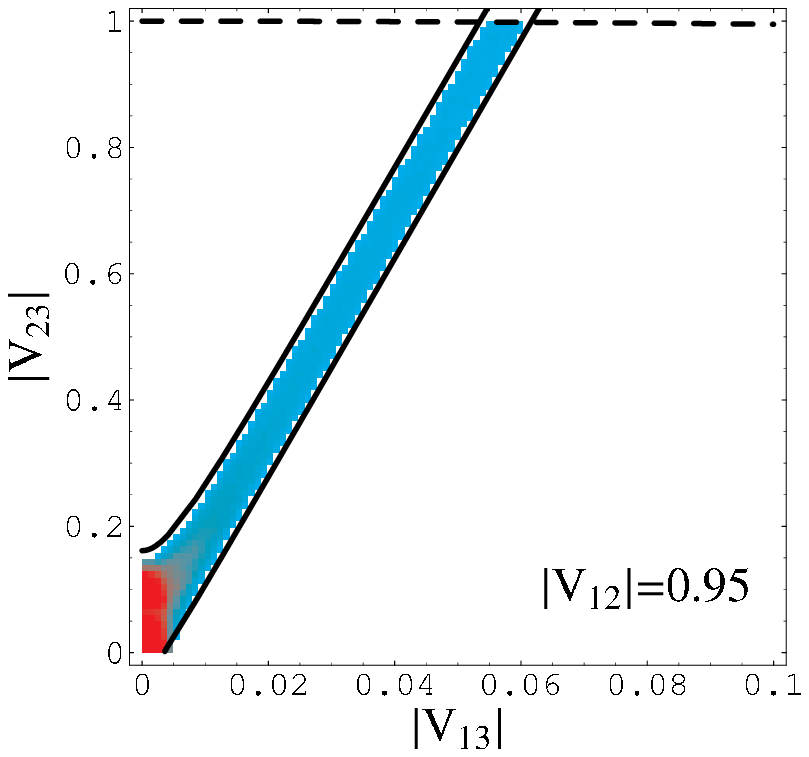,width=90mm}\\ \hspace{3.2cm}
\psfig{figure=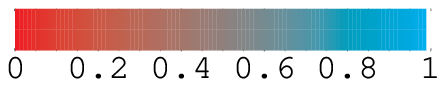,width=70mm}
\end{tabular}
\caption
{\footnotesize
Region in the $|(V_R)_{13}|-|(V_R)_{23}|$ plane which gives $m_3/m_2\leq 6$
for some choice of the phases of $V_R$ (see eq.(\ref{VR})).
For each point, 1000 random choices are probed. The
color indicates the fraction of unsuccessful choices: from 
red (complete success) to light blue.
The scenario
is defined by $M_1=M_2=M_3=y_1:y_2:y_3=1:300:9\times 10^4$.
Each plot
corresponds to a different value of $|(V_R)_{12}|$. 
The dashed line in the last plot corresponds to the limit of 
unitarity of $V_R$, while the solid line corresponds to the 
approximate analytical bound discussed at 
eq.~(\ref{haproxanalyt}). 
}
\end{figure}

The shape of the complete allowed region
can be analytically understood as
follows. For not too small $|(V_R)_{23}|^2$ [in particular when
we allow for cancellations in the numerator of (\ref{haproxp})],
the denominator of (\ref{haproxp}) is dominated by 
$|(V_R)_{31}|^2$, which satisfies the unitarity constraint
$|(V_R)_{31}|^2 \leq|(V_R)_{13}|^2 +|(V_R)_{23}|^2$.
On the other hand, the numerator of (\ref{haproxp})
is minimal when the maximum cancellation between the 
various terms occurs. Thus we can write
\bea
\label{haproxanalyt}
h\gsim
{y_3^2\over y_2^2}{M_2\over M_1}\ 
{{\rm Min} \left|\ -\left| |(V_R)_{13}|^2
\pm {M_1\over M_2}|(V_R)_{23}|^2\right| + {M_1\over M_3}\left(
1-|(V_R)_{13}|^2-|(V_R)_{23}|^2\right)\ \right|^2 
\over
|(V_R)_{13}|^2+|(V_R)_{23}|^2}
\ .
\eea
Moreover, when the two possibilities inside $|\ |^2$ in the numerator
of (\ref{haproxanalyt}) have opposite signs, then it is possible
to achieve an exact cancellation by adjusting the phases of
the various terms in the numerator of (\ref{haproxp}).
The values of $|(V_R)_{13}|$ and $|(V_R)_{23}|$ that 
saturate the approximate analytical bound (\ref{haproxanalyt}) 
for $h=6$ are indicated in the last plot of 
Fig.~3 with a solid line, which
describes the exact allowed region in a fair way.

Notice that for $|(V_R)_{23}|^2\gg |(V_R)_{13}|^2$, eq.(\ref{haproxanalyt})
gets simplified to
\bea
h\gsim \frac{y_3^2}{y_2^2} \frac{M_2}{M_1}
\frac{\left[|(V_R)_{13}|^2-\frac{M_1}{M_2} |(V_R)_{23}|^2\right]^2}
{|(V_R)_{13}|^2+|(V_R)_{23}|^2}\ ,
\label{h-3nu-approx2}
\eea
which is responsible for the long and light strip in the plots.
Notice also that for this region, the cancellation requires
the $(V_R)_{13}^2$ and $(V_R)_{23}^2$ terms in (\ref{haproxp})
to have different signs, so $\alpha_2\simeq \pm \pi/2$.

Of course, eq.(\ref{haproxanalyt})
could be further refined to include the 
effect of  $|(V_R)_{12}|$, through the modification of the
unitarity constraints on $|(V_R)_{31}|^2$, although the 
exact expression is too complicated to be of any practical
use. In any case, we already discussed the impact of the value
of  $|(V_R)_{12}|$ on the possibility to get $h^{exp}$
with no fine-tunings.

Using a less strong hierarchy for the Yukawas, such as
$y_1:y_2:y_3$ $=1:20:400$, the results are similar, except that the allowed area in Fig.~3 is larger and the required fine-tuning in the phases is less severe.

Finally note that all these results and plots apply equally for the SUSY case.

\vspace{0.3cm}
\noindent
\underline{${\bf M_1\simeq M_2\simeq M_3}$}

If $M_1= M_2= M_3$, the expression within straight brackets in the denominator
of eq.(\ref{haprox}) (which is always $\leq 1$) is naturally ${\cal O}(1)$,
unless there is some -undesired- cancellation inside. 
Hence we can write
\bea
\label{haproxpdeg3}
h={\kappa_3\over \kappa_2}\gsim
{y_3^2\over y_2^2}
\left|(V_R)_{13}^2
+ (V_R)_{23}^2+ (V_R)_{33}^2\right|^2 
\ ,
\eea
Since ${y_3^2\over y_2^2}$ is far larger than $h^{exp}\lsim 6$, a strong 
cancellation between the three terms inside the straight brackets is mandatory. 
Hence, we can already conclude that for (approximately) degenerate right-handed
masses and hierarchical Yukawa couplings (as for the other fermions), the 
observed spectrum of neutrinos can only be obtained by fine-tuning the 
high-energy parameters. 

The allowed region, $h^{exp}\leq 6$, in the 
$|(V_R)_{13}|-|(V_R)_{23}|$ plane is shown in Fig.~4 for 
fixed values of $|(V_R)_{12}|$, taking again
$y_1:y_2:y_3$ $=1:300:9\times 10^4$.
In this case the results do not depend much on the value
of $|(V_R)_{12}|$, as is clear from (\ref{haproxpdeg3}). 
The shape of the allowed region can be understood by
reasoning in a similar way as for eq.(\ref{haproxanalyt}).
Now we get
\bea
\label{haproxanalyt2}
h\gsim
{y_3^2\over y_2^2}\ 
{\rm Min} \left|\ -\left| |(V_R)_{13}|^2
\pm |(V_R)_{23}|^2\right| + \left(
1-|(V_R)_{13}|^2-|(V_R)_{23}|^2\right)\ \right|^2 
\ .
\eea
Again, when the two possibilities inside $|\ |^2$ in the numerator
of (\ref{haproxanalyt2}) have opposite signs, then it is possible
to achieve an exact cancellation by adjusting the phases of
the various terms in the r.h.s. of eq.(\ref{haproxpdeg3}). The solid
line in the first plot of Fig.~4 shows the bound $h=6$ obtained with the approximate
analytical form (\ref{haproxanalyt2}), which clearly
describes very well the exact results.

\begin{figure}
\hspace{2cm}
\begin{tabular}{c}
\psfig{figure=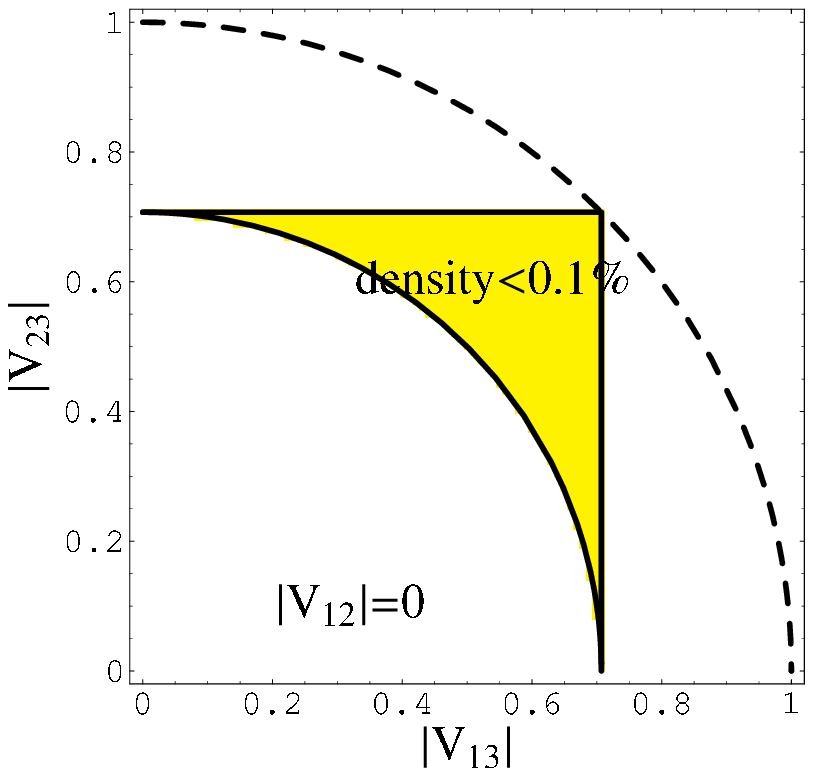,width=65mm} \hspace{0.5cm}
\psfig{figure=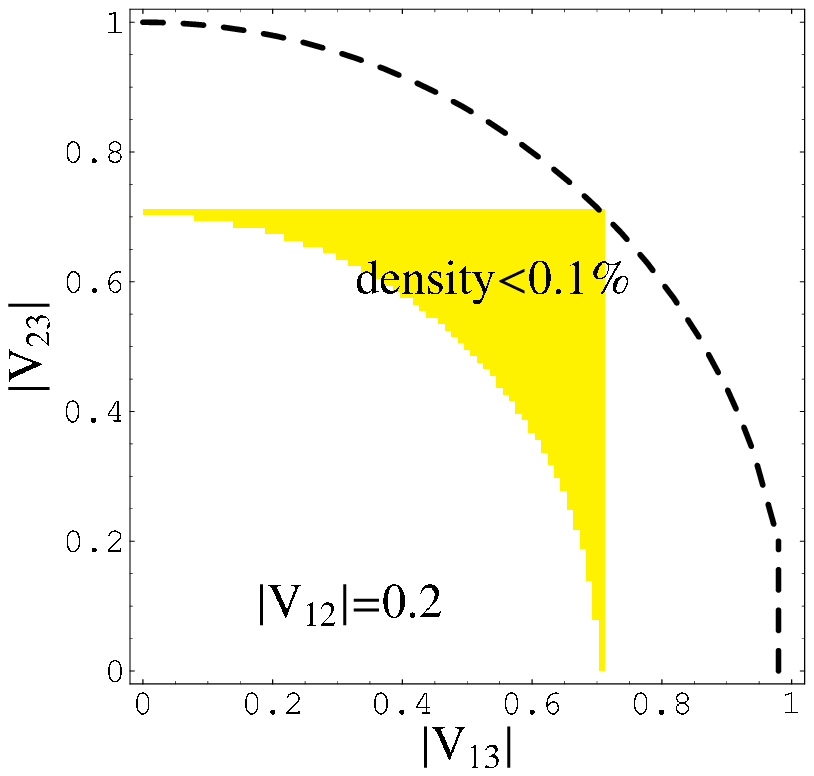,width=65mm} \\
\psfig{figure=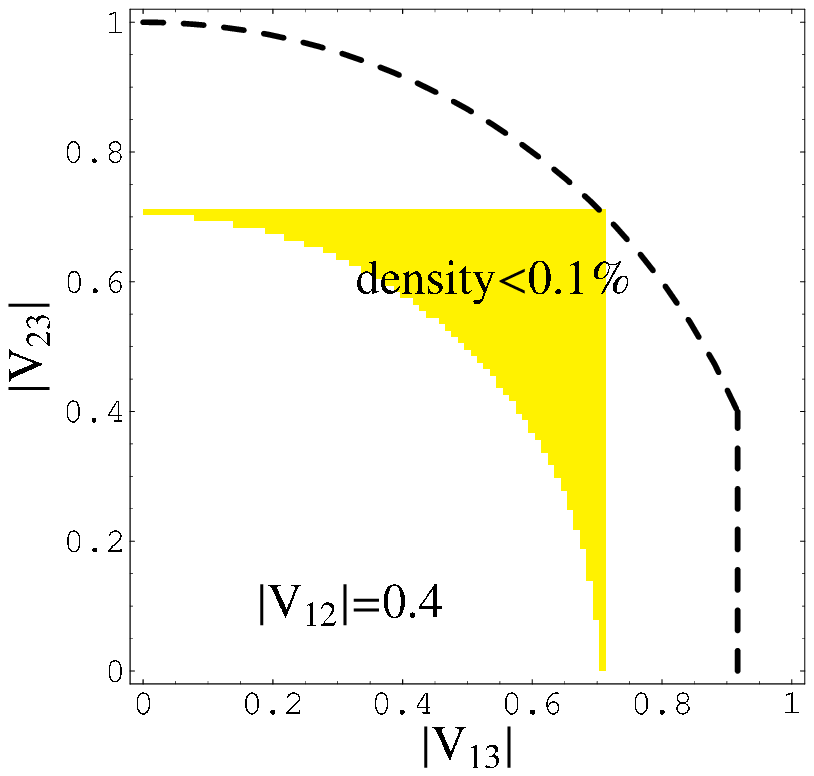,width=65mm} \hspace{0.5cm}
\psfig{figure=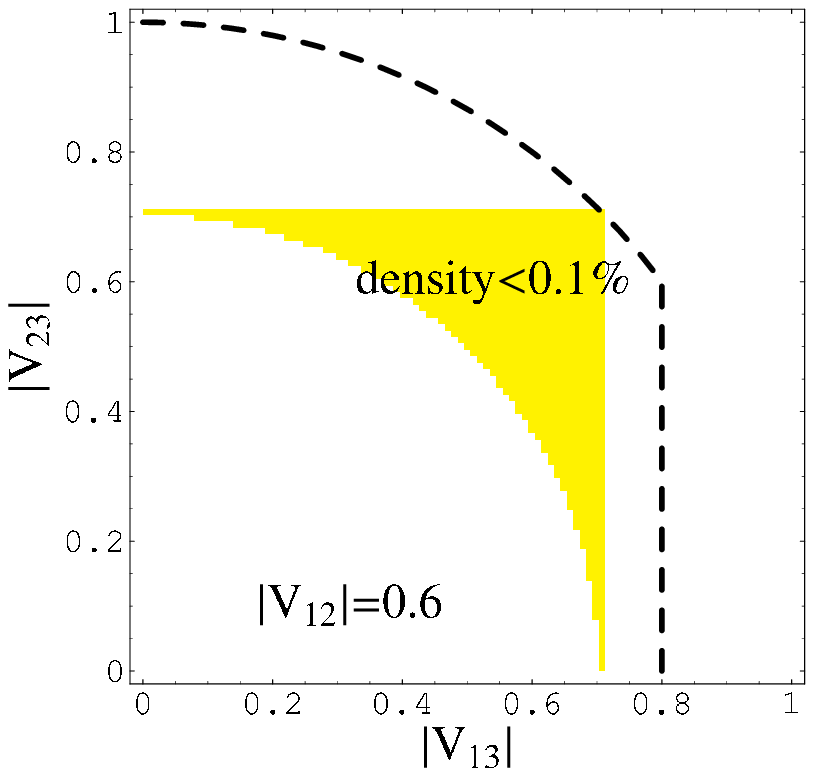,width=65mm} \\
\psfig{figure=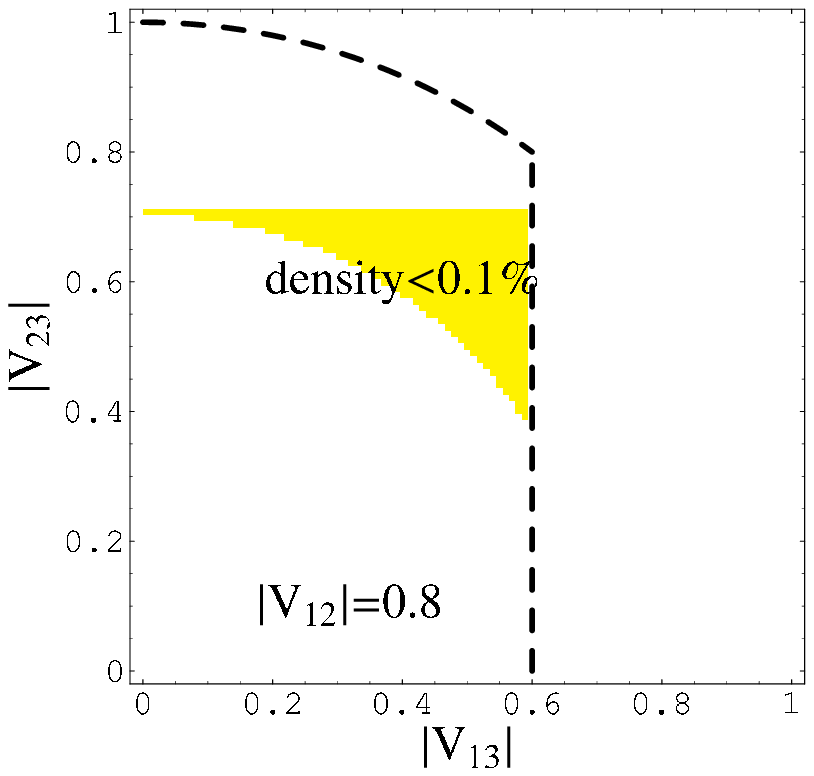,width=65mm}  \hspace{0.5cm}
\psfig{figure=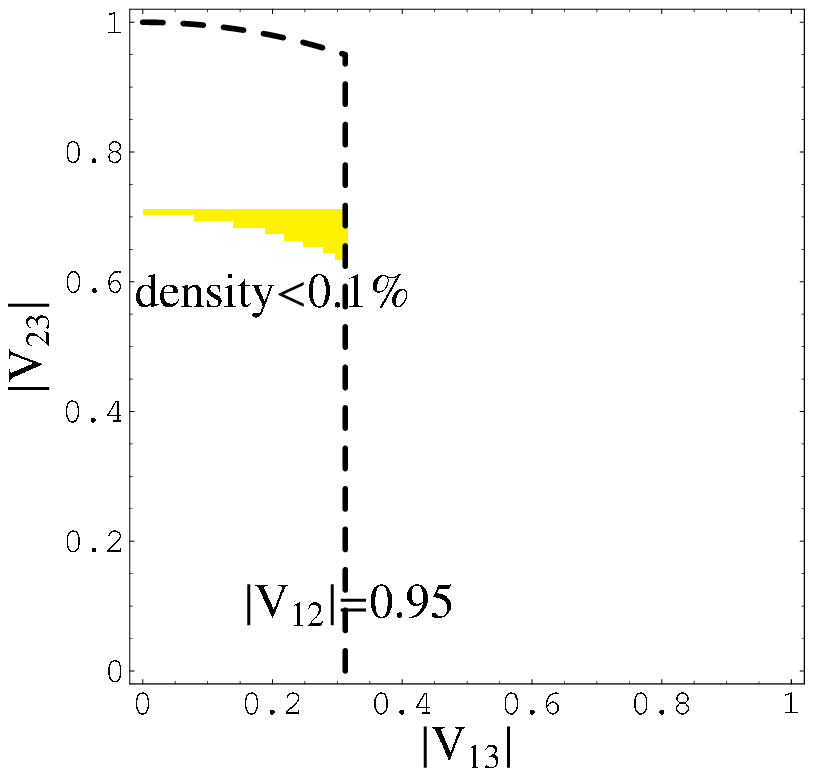,width=65mm} 
\end{tabular}
\caption
{\footnotesize  
Region in the $|(V_R)_{13}|-|(V_R)_{23}|$ plane which gives $m_3/m_2\leq 6$
for some choice of the phases of $V_R$ (see eq.(\ref{VR})). 
The yellow color indicates that in that region only less than 0.1\% of the choices
are successful.
The scenario
is defined by $M_1=M_2=M_3$ and $y_1:y_2:y_3=1:300:9\times 10^4$.
Each plot
corresponds to a different value of $|(V_R)_{12}|$. 
The dashed line
is the limit of unitarity of $V_R$, while the solid line in the first
plot corresponds to the approximate analytical bound discussed at 
eq.(\ref{haproxanalyt2}).}
\end{figure}


It is worth mentioning that in this case a CKM-like form for $V_R$
cannot lead to a realistic spectrum, since [for any choice of
the phases in eq.(\ref{VR})] it is not consistent with a cancellation
in the r.h.s. of eq.(\ref{haproxpdeg3}). However, it is funny that
a MNS-like form can work correctly. More precisely, when $|(V_R)_{13}|\ll 1$,
as is the MNS case, the condition for cancellation in  eq.(\ref{haproxpdeg3})
is approximately $|(V_R)_{13}|^2\pm (|(V_R)_{23}|^2-|(V_R)_{33}|^2) \simeq 0$.
In terms of the parametrization (\ref{VR}), this reads
\bea
\label{condVRdeg}
\tan^2\theta_{13}^R \simeq |\cos 2\theta_{23}^R|
\ .
\eea
This condition is precisely fulfilled by an MNS-like matrix, thanks to
the smallness of $\theta_{13}$ and the near-to-maximal $\theta_{23}$.

\section{A suggestive ansatz}

In sect. 4 we have not made any particular assumption about the (high-energy) 
parameters of the see-saw, apart from considering hierarchical neutrino
Yukawa couplings, similar to those of quarks and charged leptons. Nevertheless,
we showed that if the right-handed neutrino masses are hierarchical, a CKM-pattern
for $V_R$ was naturally preferred in order to reproduce the experimental
ratio between the two heavier neutrinos, 
$h={\kappa_3/ \kappa_2}\lsim 6$, which is 
the only experimental constraint on ratios of neutrino masses.
Similarly, we saw that if the right-handed neutrino masses are approximately
degenerate, an MNS-like pattern for $V_R$ could equally work, but always with
a certain fine-tuning. In this section we study more in deep these suggestive 
coincidences.

\vspace{0.3cm}
\noindent
\underline{${\bf V_R=V_{\rm {\bf CKM}}}$ {\bf ansatz}}

\vspace{0.2cm}
\noindent
We start 
by considering the possibility that $V_R$ coincides with the CKM matrix, 
$V_{\rm CKM}$.
From eq.(\ref{VR}) $V_R$ has two phases, $\alpha_1, \alpha_2$, that, unlike
the quark CKM matrix, cannot be absorbed into redefinitions of the 
fields. Thus, the identification of $V_R$ with $V_{\rm CKM}$ has to be
up to these two independent phases,
\bea
\label{VRVCKM}
V_R={\rm diag}\left(e^{i\alpha_1},\ e^{i\alpha_2},\ 1\right)V_{\rm CKM}
\eea
This identification of $V_R$ with $V_{\rm CKM}$ evokes the
$SU(5)$ connection between the $V_L$ mixing matrix for $d-$quarks and 
the $V_R$ one for
charged leptons, which comes from the relation ${\bf Y}_d={\bf Y}_l^T$
between the corresponding Yukawa matrices.
Following this analogy, we can make the ansatz that the eigenvalues
of neutrino Yukawa couplings, $\{y_1, y_2, y_3\}$, coincide with
the $u-$quark ones, $\{y_u, y_c, y_t\}$. We are {\em not} considering a 
definite GUT framework to justify this assumption (although it could 
proceed e.g. from some $SO(10)$ construction), but only exploring
if it can work in practice, which is certainly non-trivial.

The first step to probe this ansatz is to write both $V_{\rm CKM}$ and
$\{y_u, y_c, y_t\}$ at the scale of right-handed masses, $M\sim 10^{13}$
GeV, where the see-saw mechanism takes place and the identification
(\ref{VRVCKM}) should be done\footnote{A more GUT-inspired alternative 
is to run $V_{\rm CKM}$ up to $M_X$, perform the 
identification (\ref{VRVCKM}) and then run $V_R$ down to the seesaw scale.
This procedure is more cumbersome and, given the closeness of the 
$M$ and $M_X$ scales, the former approach is sufficiently
precise.}. 
In the SM the RG change in the ratios $m_u:m_c:m_t = y_u:y_c:y_t$
from low- to high-energy is
\bea
 y_u: y_c: y_t\ &=&\ 1.3\times 10^{-5} : 7.1\times 10^{-3}: 1
\;\;\;{\rm at\ low\ scale}\nonumber\\ 
\rightarrow\ 
 y_u: y_c: y_t\ &=&\ 1.1\times 10^{-5} : 3.2\times 10^{-3}: 1
\;\;\;{\rm at\ high\ scale}\ 
\label{yuRG}
\eea
Note that the RGE change considerably the hierarchy of $u-$quarks
(which, incidentally, becomes remarkably regular, on top of strong ).
This is due mainly to the important effect
of the top Yukawa coupling. On the other hand, the RGE for the neutrino mass
matrix below the $M-$scale is flavour-blind, except for small effects 
proportional to the squared of the tau Yukawa coupling. This produces
very small effects in the hierarchy of neutrino masses and in the
MNS matrix (which we are not considering here anyhow),
especially in the case of a soft hierarchy \cite{Casas:1999tg}.
Thus we can neglect here the RGE effects for the neutrino sector.
$V_{\rm CKM}$ undergoes a certain change as well for the same reasons. 
In magnitude,
\bea
\label{VCKM}
\left|V_{\rm CKM}\right|\simeq
\pmatrix{0.97&0.23&0.0043\cr&&\cr 0.23& 0.973& 0.042
\cr&&\cr 0.008&0.04&1\cr}_{\rm low\ scale}
\hspace{-0.5cm}\longrightarrow\ \ 
\pmatrix{0.97&0.23&0.0049\cr&&\cr 0.23& 0.973& 0.047
\cr&&\cr 0.009&0.047&1\cr}_{\rm high\ scale}
\eea
The CP-phase, $\delta_{\rm CKM}\simeq 1$ rad, does not change appreciably
along the running. Of course, eqs.~(\ref{yuRG}, \ref{VCKM}) have experimental
errors. For our purposes the most significant ones are those associated
to $(V_{\rm CKM})_{13}$ and $(V_{\rm CKM})_{23}$. Using the 
most recent analyses \cite{Yao:2006px}
and running consistently the quoted errors up to the $M-$scale 
\cite{Olechowski:1990bh,Ramond:1993kv},
we get $(V_{\rm CKM})_{13}=(4.9\pm 0.3)\times 10^{-3}$, 
$(V_{\rm CKM})_{23}=(47\pm 0.7)\times 10^{-3}$.

In addition we will consider, as mentioned,  hierarchical right-handed 
masses, choosing a hierarchy equal to that of the Yukawa couplings. 
This is
of course a somewhat arbitrary choice, but we find it simple and
reasonable, and
it does not amount to any extra assumption for a different hierarchy.

In summary, we will make the assumption
\bea
\label{hierarchyM}
y_1:y_2:y_3&=&M_1:M_2:M_3\ =\ 1.1\times 10^{-5} : 3.2\times 10^{-3}: 1
\nonumber\\
V_R&=&{\rm diag}\left(e^{i\alpha_1},\ e^{i\alpha_2},\ 1\right)V_{\rm CKM}(M)
\eea
where $V_{\rm CKM}(M)$ is essentially given by (\ref{VCKM}).

Notice from (\ref{seesaw3}, \ref{seesaw2}) that choosing 
$V_R={\bf 1}$ we would get a hierarchy of neutrino masses equal to that of
Yukawa couplings, i.e. $h=\kappa_3/\kappa_2\sim 300$ [see
eq.(\ref{hierarchyM})]. This would be 
completely inconsistent with the experimental 
$h=\kappa_3/\kappa_2\lsim 6$,
by a factor of 50. 
On the other hand, as is clear from the discussion 
around eq.(\ref{haprox}), 
a random $V_R$ would give $h = {\cal O}(10^7)$, i.e.
orders of magnitude away from the experimental range.
Therefore,
it is certainly non-trivial
that the assumption (\ref{hierarchyM}) could be 
consistent with the experiment.

To illustrate these facts and show the results, we give 
in Fig.5, upper plots, the allowed region 
in the $|(V_R)_{13}|-|(V_R)_{23}|$
plane for fixed $|(V_R)_{12}|= |(V_{\rm CKM})_{12}|$. 
Again, for each point we have evaluated
$h=\kappa_3/\kappa_2$ for 1000 random values of the $\alpha_1$, $\alpha_2$
phases in $V_R$ ($\delta_R$ is fixed at $\delta_{\rm CKM}$), and counted the number of points that are compatible
with the observed hierarchy, $h\lsim 6$. White areas are excluded, while
colored areas are allowed. As expected only a tiny part of 
$\{|(V_R)_{13}|, |(V_R)_{23}|\}$ values are allowed [a good analytical
approximation of the size and shape of the allowed region is given by 
(\ref{haproxanalyt})]. 
Remarkably,
the CKM value for these quantities (represented by the cross
in the figure) falls inside the allowed region, which
we find very suggestive and highly non-trivial. Notice
also that $V_{\rm CKM}$ is the only experimentally known example of
a mixing matrix for Yukawas\footnote{Recall that, if desired, one can go
to a basis
of quark doublets where $V_{\rm CKM}$ is associated just to ${\bf Y_d}$ 
or ${\bf Y_u}$.}, as $V_R$ is ($V_{\rm MNS}$ is not, unless
neutrinos are pure Dirac). 
All this makes the success of the CKM 
ansatz even more remarkable. It would be certainly nice
to construct models (maybe in the GUT framework) to
accommodate this ``CKM-ansatz''.

\begin{figure}
\hspace{-2.5cm}
\begin{tabular}{c}
\psfig{figure=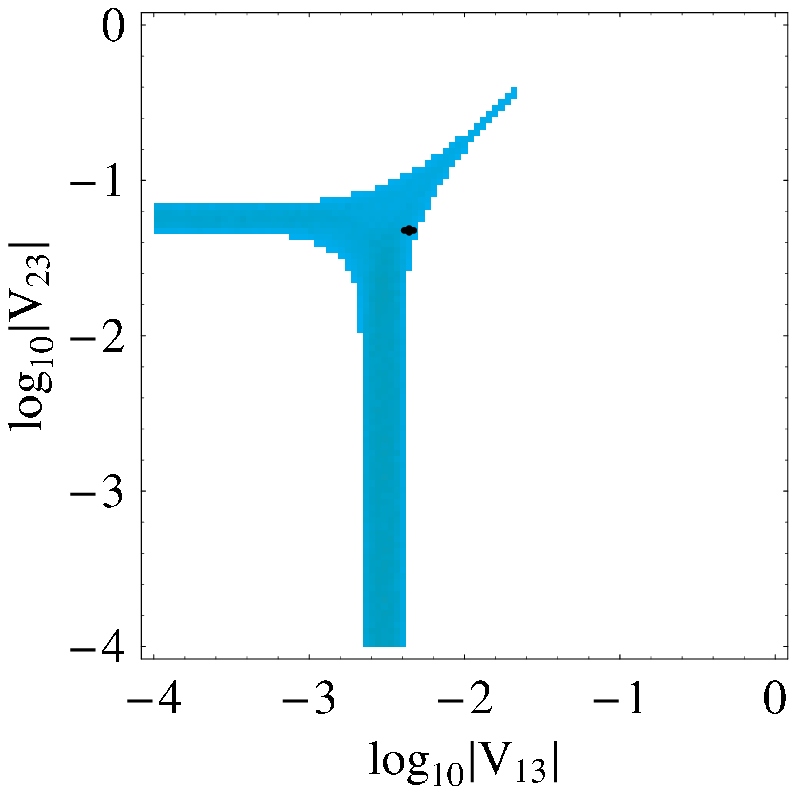,width=90mm} \hspace{-2cm}
\psfig{figure=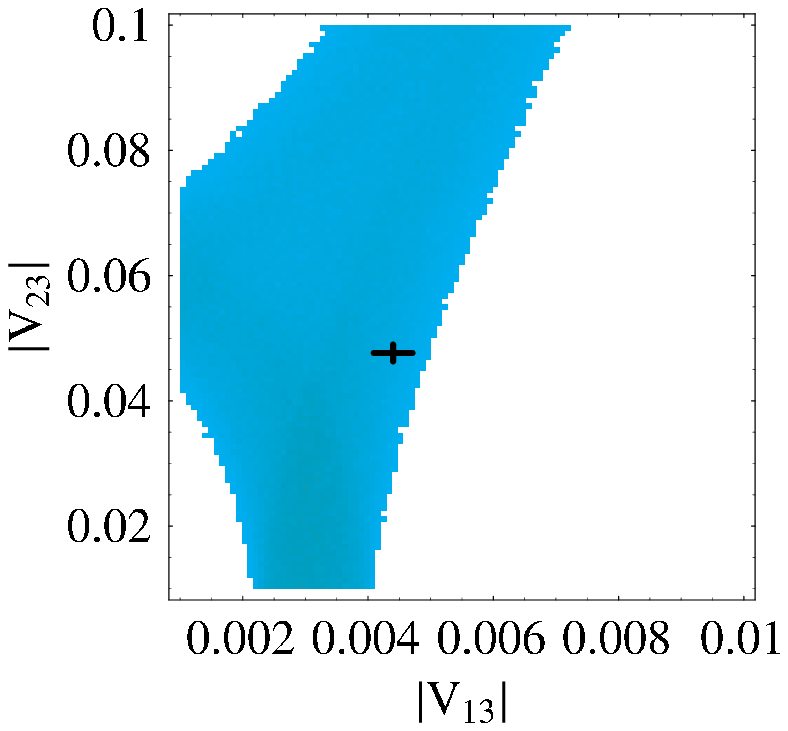,width=90mm} \\
\psfig{figure=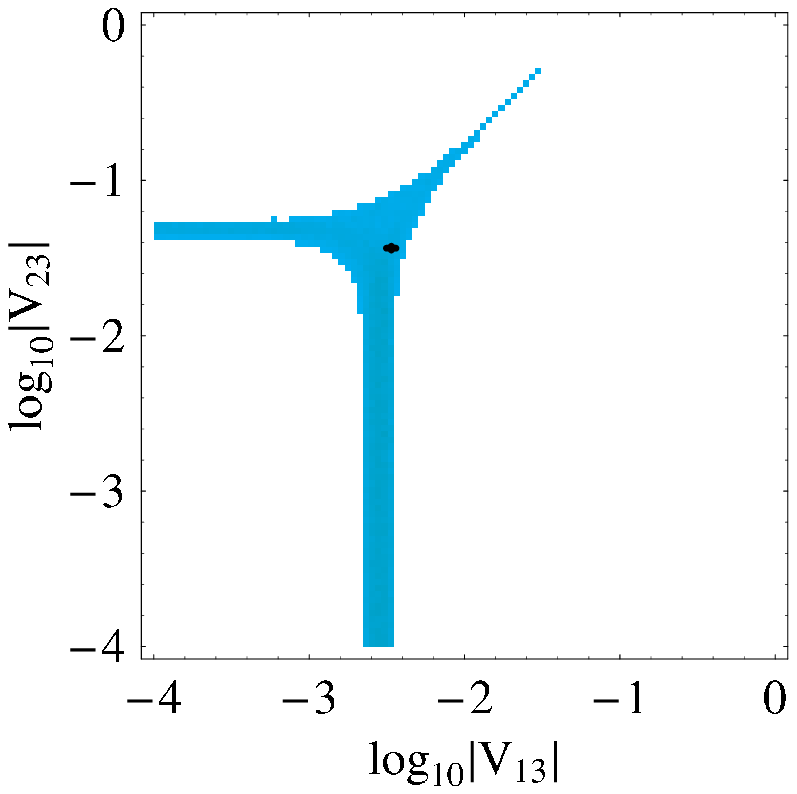,width=90mm} \hspace{-2cm}
\psfig{figure=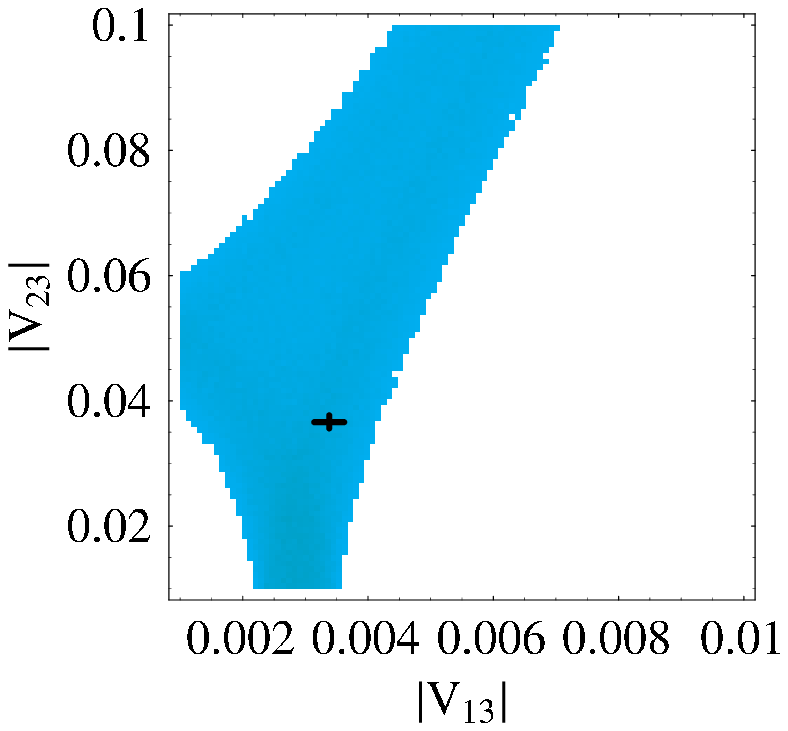,width=90mm} 
\end{tabular}
\caption
{\footnotesize 
Region in the $|(V_R)_{13}|-|(V_R)_{23}|$ plane which, 
for $|(V_R)_{12}|=|(V_{\rm CKM})_{12}|$, $\delta_R = \delta_{\rm CKM}$,
and some 
choice of the $\alpha_1$, $\alpha_2$ phases (see eq.(\ref{VR})),  
gives $m_3/m_2\leq 6$. The Yukawa couplings, $y_i$, and right-handed masses, $M_i$,
are taken as indicated at eq.(\ref{hierarchyM}). The cross corresponds to the
CKM values for $|(V_R)_{13}|$, $|(V_R)_{23}|$ (within experimental uncertainties).
The upper (lower) plots correspond to the SM (SUSY) case. The color code is as in
Fig.~3.
}
\end{figure}


In order to gain analytical understanding for the success of the ``CKM-ansatz''
it is convenient to use the Wolfenstein parametrization of the CKM matrix:
\bea
V_{CKM}= \pmatrix{1-{\displaystyle{\lambda^2\over 2}} & \lambda &
A\lambda^3(\rho-i\eta)\cr
&&\cr
-\lambda & 1- {\displaystyle{\lambda^2\over 2}} & A\lambda^2 \cr
&&\cr
A\lambda^3(1-\rho-i\eta) & -A\lambda^2 & 1 \cr} + {\cal O}(\lambda^4) ,
\label{Wolfenstein}
\eea
where $\lambda$ is determined with a very good precision in
semileptonic $K$ decays, giving $\lambda \simeq 0.23$, and $A$ is measured
in semileptonic $B$ decays, giving $A\simeq 0.82$. The parameters
$\rho$ and $\eta$ are more poorly measured, although a rough
estimate is $\rho\simeq 0.1$, $\eta\simeq 0.3$ \cite{Battaglia:2003in} 
(therefore  $(\rho-i\eta)\simeq 0.3 e^{i\delta}$, which is fairly close to
$\lambda$ in absolute value).
At high energies, only the parameter $A$ changes substantially \cite{Ramond:1993kv},
being $A\simeq 0.92$ at the scale $M\sim 10^{13}$GeV.
Furthermore, we will use the following phenomenological relations among the 
up-type quark Yukawa couplings evaluated at high energies, that we assume 
also valid for the right-handed neutrino masses:
\bea
y_1:y_2:y_3&\sim&M_1:M_2:M_3\ \sim \ \lambda^8 : \lambda^4: 1
\eea
Substituting this ansatz in eq.~(\ref{haprox}) we obtain:
\bea
\label{haproxCKM}
h= {m_3\over m_2}\ \gsim \ \lambda^{-6}\ \frac
{\left|  A^2 (\rho-i\eta)^2 e^{2i\alpha_1}+ A^2\lambda^2 e^{2i\alpha_2}+ \lambda^2\right|^2}
{\left|A^2(1-\rho-i\eta)^2+1+ \lambda^2 e^{2i\alpha_2}\right|}\sim {\cal O}(\lambda^{-2})
\eea
It is already remarkable the large reduction of the hierarchy that results
just from the peculiar pattern of $V_{\rm CKM}$ (without taking into account
the values of  $\alpha_1$, $\alpha_2$): for random $V_R$, the natural size 
of the hierarchy is dictated by the
 ${y_3^2\over y_2^2}{M_2\over M_1}\sim \lambda^{-12}$ factor in  eq.~(\ref{haprox}).
Now, thanks to the structure of $V_{\rm CKM}$
given in eq.(\ref{Wolfenstein}), the second factor in eq.~(\ref{haprox}) (i.e. the
fraction of absolute values) gets ${\cal O}(\lambda^{10})$, leading to (\ref{haproxCKM}).
Plugging numbers, for random  $\alpha_1$, $\alpha_2$, this 
amounts to a reduction from
$h\sim{\cal O}(10^7)$ to $h\sim 100$.  
This is still too large compared to 
$h^{\rm exp}\sim 6$, but shows that $V_{\rm CKM}$ does soften $h$ in an extremely
efficient way. Choosing $\alpha_1 - \delta \sim {\pi\over 2}, {3\pi\over 2}$ and
$\alpha_2 \sim 0,\pi$ the numerator of eq.~(\ref{haproxCKM}) 
gets much smaller due to a cancellation among the three terms. This is possible
thanks to the fact that the three terms have similar magnitude, which is a 
fortunate coincidence (changing $V_R$, even keeping the same pattern,
this fact generally disappears). Then we get $h={\cal O}(1)$, i.e. consistent
with the experiment. This choice of phases is as good as any other else,
implying that there is no need of {\em fine} tuning of the phases to get
the desired result.

Coming back to the numerical computation, the previous arguments are
illustrated in 
Fig.6, left plot, which shows the region of experimentally acceptable
values of $h$ in the $\alpha_1-\alpha_2$ plane. More precisely, the 
green area corresponds to $5.5\leq h\leq 6$, which is the experimental
$1-\sigma$ value of $h^{exp}$ when $m_2/m_1\gg 1$  (see
Fig.~1), as is the case. As noted above this allowed region replicates
with periodicity $\pi$. 
All the remaining parameters of $V_R$ have been taken at the central values
of $V_{\rm CKM}$. Clearly, the allowed region for
$\alpha_1$, $\alpha_2$ is quite ``macroscopic", i.e. it is
not fine-tuned. In fact, the minimal value for $h$ is close
to the experimental value $h\sim 6$ (note that since $\kappa_1$
is hierarchically smaller, as will be commented shortly, the value of $h$
must be close to its experimental upper bound). This is 
funny since the region of minimal values of $h$ is naturally
enhanced in size (near a minimum the function changes little).

\begin{figure}
\begin{tabular}{c}
\psfig{figure=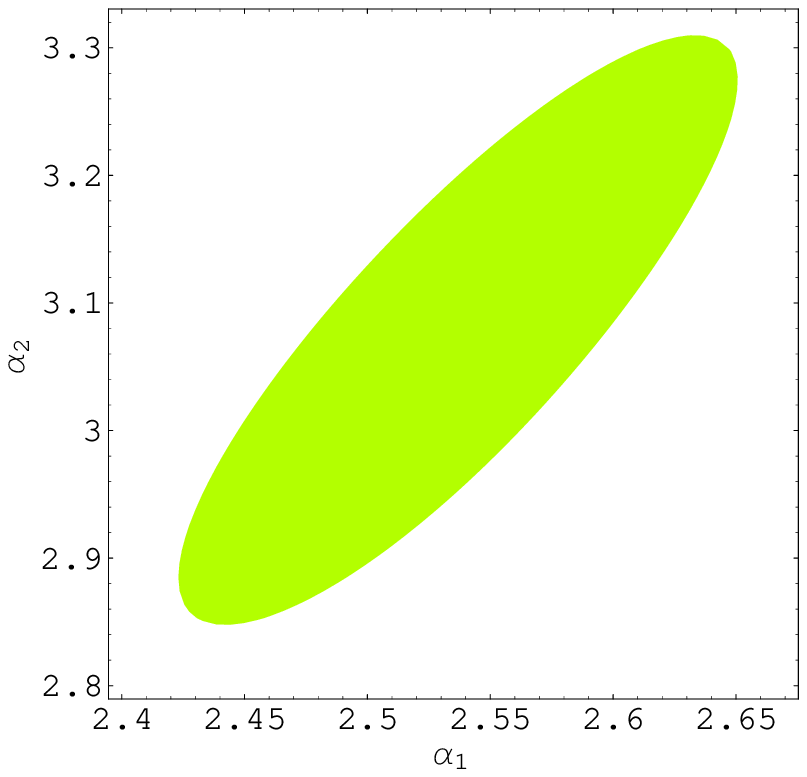,width=75mm}
\psfig{figure=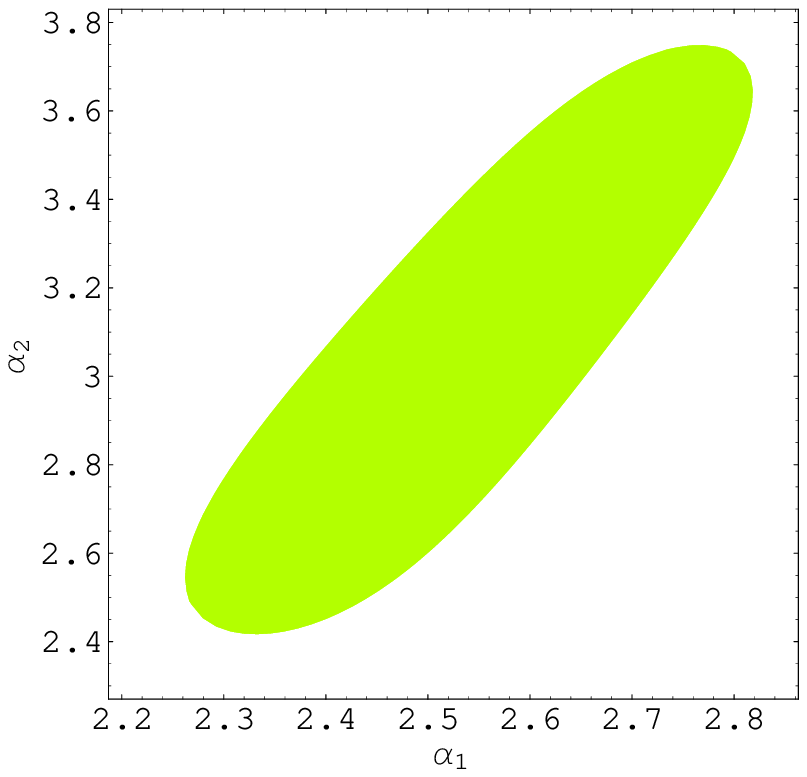,width=75mm}
\end{tabular}
\caption
{\footnotesize
Region of the $\alpha_1-\alpha_2$ plane that gives values
of $m_3/m_2$ consistent with the experiment for all the remaining parameters of $V_R$ (see 
Right
and left plots correspond to
the SM and SUSY cases respectively. Note the different scales of the two plots.
}
\end{figure}


Let us indicate that the mass of the lightest neutrino, 
$\kappa_1$, becomes orders of magnitude smaller than $\kappa_2$,
in agreement with the general results of sect. 4.3 (see the
discussion after eq.[\ref{haproxp})]. To be precise, the
value of the lightest neutrino mass predicted by this ansatz is
\bea
\label{m1tiny}
m_1=v \kappa_1 \simeq 3\times 10^{-6} m_2 = 3\times10^{-8}{\rm eV}
\eea

The SUSY case works in a similar way. The main difference are the 
RGEs, which are a bit different and, besides, depend on the value
of $\tan\beta$, though not dramatically. 
The results
for the CKM ansatz are also similar, and even better, 
as shown in Fig.~5 (lower plots) and Fig.~6 (right plot) 
for a typical case ($\tan\beta = 10$).

Finally, let us mention that choosing a hierarchy for the Yukawa couplings
as that of $d-$quarks (which is quite milder) enhances the allowed region
in the $|(V_R)_{13}|-|(V_R)_{23}|$ plane. Then the CKM point continues to fall 
inside the allowed region.

\vspace{0.3cm}
\noindent
\underline{${\bf V_R=V_{\rm {\bf MNS}}}$ {\bf ansatz}}

\vspace{0.2cm}
\noindent
Let us now consider the $V_R\sim V_{\rm {\bf MNS}}$ possibility. As
discussed at the end of subsect.4.3, this can work if the right-handed masses
are quasi-degenerate; for simplicity we will assume $M_1=M_2=M_3$.
As for the CKM case, the identification of $V_{\rm {\bf MNS}}$ 
and $V_R$ can only be made up to the two independent $\alpha_1$,
$\alpha_2$ phases in (\ref{VR}). The Majorana phases of 
$V_{\rm {\bf MNS}}$ act from the opposite side, see eq.(\ref{UV}),
and cannot be identified with $\alpha_1$, $\alpha_2$. In any case,
we do not have any experimental information about these Majorana phases,
nor about $\delta$, in the MNS matrix. So we take
\bea
\label{VRVMNS}
V_R={\rm diag}\left(e^{i\alpha_1},\ e^{i\alpha_2},\ 1\right)V
\eea
where $V$ is the ``non-Majorana" part of the MNS matrix, given in eq.(\ref{Vdef}).
More precisely \cite{Maltoni:2004ei},
\bea
\label{angles}
\sin^2\theta_{12}=0.26-0.36,\;\;\;\;
\sin^2\theta_{23}=0.38-0.63,\;\;\;\;
\sin^2\theta_{13}\leq 0.025,
\eea
the value of $\delta$ is left free. 
Concerning Yukawa couplings, as in the CKM case we identify them with the $u$--quark
Yukawa couplings at high energy, which for the SM are given in eq.(\ref{yuRG}).

The results are given in Fig.7, left plot, which shows the allowed region 
in the $|(V_R)_{13}|-|(V_R)_{23}|$
plane for fixed $|(V_R)_{12}|= |(V_{\rm MNS})_{12}|$.
Again, colored areas are consistent (for some choice of the phases) with 
the observed hierarchy, $h\lsim 6$, while white areas are excluded.
The MNS value for $|(V_R)_{13}|,\ |(V_R)_{23}|$ is represented by a cross
in the figure, falling inside the allowed region. 

\begin{figure}
\hspace{-2.5cm}
\begin{tabular}{c}
\psfig{figure=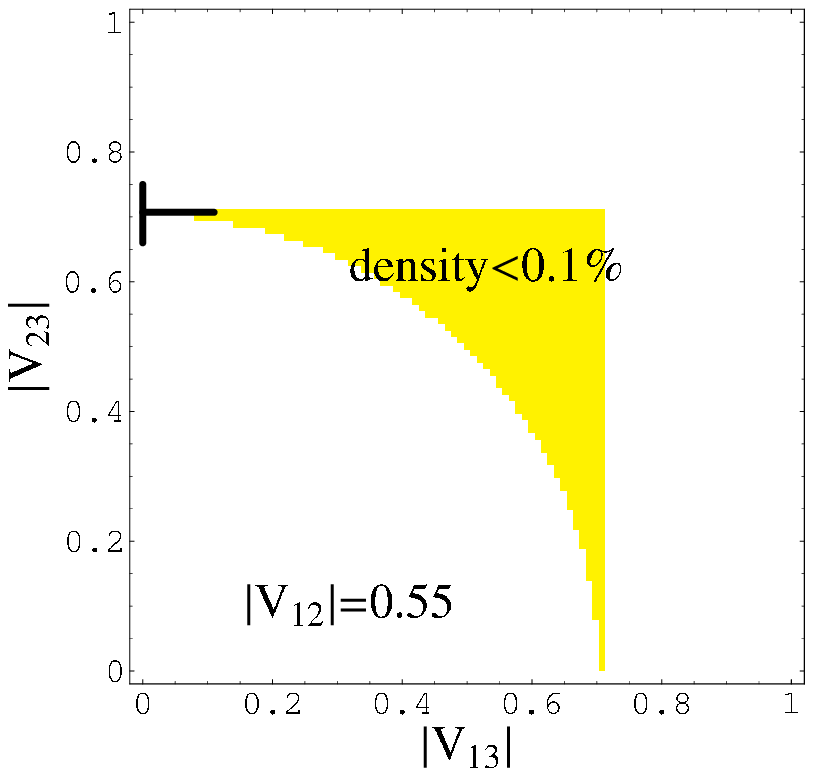,width=90mm} \hspace{-2cm}
\psfig{figure=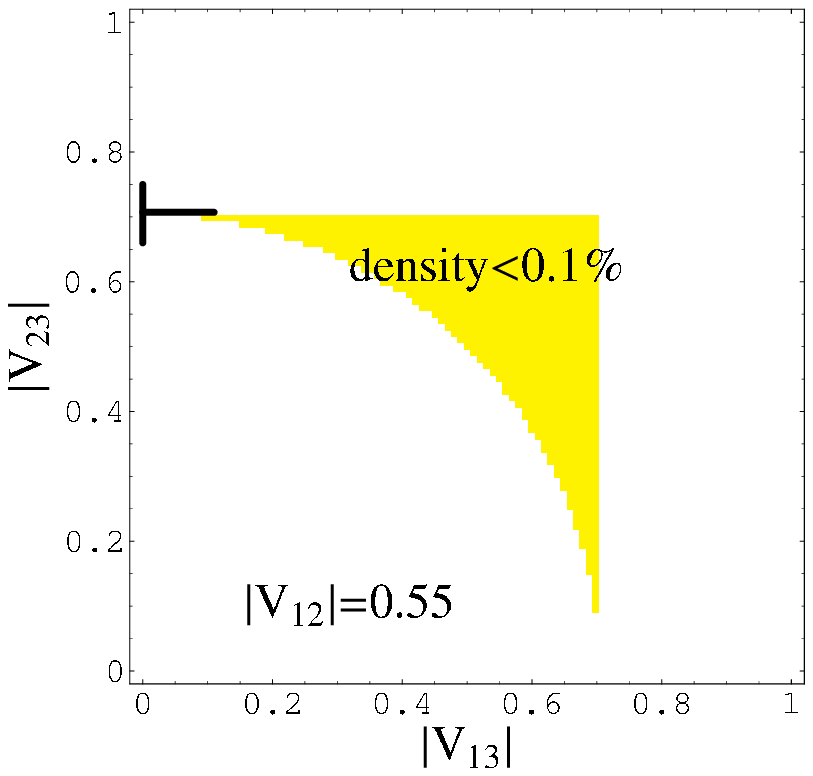,width=90mm}
\end{tabular}
\caption
{\footnotesize  
The same as Fig.~5 but for $V_{\rm CKM}\rightarrow V_{\rm MNS}$
and taking $M_1=M_2=M_3$. The $\delta_{\rm MNS}$ phase is left free, since
it is experimentally unknown. The left (right) plot corresponds to the SM (SUSY)
case. The color code is as in Fig.~4.}
\end{figure}


Although this is perhaps
less suggestive than the good performance of $V_{\rm CKM}$ in the case
of hierarchical right-handed masses, it is still quite remarkable.
Concerning the values of the phases that do the job, it is clear from
(\ref{haproxpdeg3}) that the necessary cancellation inside the straight brackets 
requires in this case $\alpha_2 \simeq \pm i \pi$, since $(V_R)_{13}^2\simeq 0$.
The previous cancellation must be quite fine as can be seen noting that
the ratio of squared Yukawa couplings in the right hand side of (\ref{haproxpdeg3})
is $\sim 10^{5}$, so the $|\ |^2$ factor must be very small in order to
obtain $h\simeq 6$ (a tunning of $\lsim 1\%$ is needed).

The performance of the SUSY case is similar, as shown in Fig.~7, right plot.

\section{Summary and conclusions}

In this paper we have started from
the fact that
the observed mass ratio for the two heavier low-energy neutrinos,
$h=m_3/m_2\lsim 6$, is much smaller than the corresponding ratios
observed for quarks and charged leptons, which are 
${\cal O}(20)$  or ${\cal O}(300)$
(for the other independent neutrino mass ratio, 
$m_2/m_1$, there is no experimental constraint).
We have wondered whether this peculiar pattern of
neutrino masses can be a consequence of the peculiar way they are generated
through a see-saw mechanism, investigating 
how the present experimental data restrict 
the structure
of the high-energy seesaw parameters and which choices, among
the allowed ones, produce more naturally
the observed pattern of neutrino masses.
We have studied in particular (but not only) if starting with hierarchical neutrino
Yukawa couplings, as for the other fermions, one can naturally get
the observed $m_3/m_2\lsim 6$ ratio.

To perform this analysis we have first put forward a top-down parametrization
of the see-saw mechanism in terms of (high-energy) basis-independent quantities:
the Yukawa and right-handed-mass "eigenvalues", $\{y_i, M_i\}$, and two
unitary matrices, $V_L, V_R$, associated to the diagonalization of the
Yukawa matrix, as shown
in eqs.(\ref{DiagM}, \ref{DiagY}). From these 18 independent parameters,
we have shown that the neutrino mass eigenvalues
depend just on 12 of them: $\{y_i, M_i\}$ and $V_R$, which simplifies the analysis a lot.
On the other hand, $V_L$ can be derived from the other parameters and 
$V_{\rm MNS}$. 
This is summarized in eqs.(\ref{list}, \ref{seesaw2}--\ref{VLW}). A parametrization
of $V_R$ is given in (\ref{VR}).

In our analysis (which is valid for both the SM and the SUSY versions of the 
see-saw)
we have made an extensive use of some analytical inequalities satisfied
by the eigenvalues
of a general hermitian matrix. This allows to obtain very simple expressions that describe
faithfully the exact results and permit to gain intuition on the problem, e.g.
the useful lower bound on $h$ given by eq.(\ref{haprox}).
This analytical study was complemented by a numerical and statistical survey,
in order to obtain and present accurate results.

Our main conclusions are the following:

\begin{itemize}

\item 

For random values of the
$V_R$ entries we expect a low-energy  neutrino hierarchy
$h={m_3\over m_2} \gsim \ 
{y_3^2\over y_2^2}{M_2\over M_1}$.
If the Yukawa couplings are hierarchical, similarly to the other fermions,  
then we expect $h$ orders of magnitude
larger than the experimental value and the hierarchy of Yukawas itself.
So, 
either we give up the natural assumption
that the neutrino Yukawa couplings present a hierarchy 
similar to other fermions, or we accept that the $V_R$ 
entries are far from random. In the second case the structure
of $V_R$ becomes strongly constrained. In particular, from 
eq.(\ref{haprox}), $\left|(V_R)_{13}^2
+ {M_1\over M_2}(V_R)_{23}^2+ {M_1\over M_3}(V_R)_{33}^2\right|^2\ll 1$ 
is required, and $(V_R)_{12}$ sizeable is desirable.

\item 

If we keep the assumption of hierarchical neutrino Yukawa couplings,
a low-energy spectrum of quasi-degeneracy or soft hierarchy
for the {\em three} neutrinos requires either
$M_i/M_j \simeq y_i^2/y_j^2$, $V_R\simeq {\bf 1}$, or a very delicate 
tuning between $\{y_i, M_i\}$ and $V_R$. In the absence of an explanation for
this strong fine-tuning we consider this scenario as unnatural. 

\item 

On the other hand, if we just attempt to reproduce the only experimentally
constrained mass ratio, $h=m_3/m_2\lsim 6$, the prospects are much more interesting:
a characteristic pattern for the $V_R$ matrix emerges, but there is no need
of fine-tuning between the parameters.

 \begin{itemize}

 \item
 If the right-handed neutrino masses are hierarchical,
 $M_1\ll M_2\ll M_3$, the selected pattern for $V_R$ 
 is characterized by very small 
 $|(V_R)_{13}|$, $|(V_R)_{23}|$, and sizeable $|(V_R)_{21}|$,
 which remarkably resembles the structure of the CKM matrix.
 (actually the discussion before eq.(\ref{haproxp2}) suggests
 $|(V_R)_{13}|^2\ll |(V_R)_{23}|^2$, also in coincidence with
 CKM).

 \item
 If the right-handed neutrino masses are degenerate,
 $M_1\simeq M_2\simeq M_3$, it is not possible to
 reproduce $h^{\rm exp}$ without a certain fine-tuning.
 The selected form for $V_R$ is not compatible with 
 $V_{\rm CKM}$, but, quite amusingly, it is with $V_{\rm MNS}$
 (altough, in this case, other patterns for $V_R$ very different
 from $V_{\rm MNS}$ work as well).

 \end{itemize}
 
In all the cases, the mass of the lightest neutrino, $m_1$, is naturally
orders of magnitude smaller than $m_2$, which comes out as a natural
prediction of a scenario with hierarchical neutrino Yukawa couplings.

\item
Motivated by the previous coincidences we have explicitely checked
that identifying $V_R$ with $V_{\rm CKM}$ and taking a hierarchy
of neutrino Yukawa couplings (and right-handed masses) equal to that
of the $u-$quarks, gives $h$ consistent with the experimental limit,
$h^{\rm exp}\lsim 6$.
This is highly non-trivial since $V_R={\bf 1}$ gives $h\simeq 300$
and a random $V_R$ typically
gives $h={\cal O}(10^6)$. 
We have not attempted to construct a GUT model to accommodate 
this suggestive feature, but it might be an interesting line of work.
For the SUSY case there are slight differences coming from the form of the RGE, but the results are very similar (and even better).

Likewise using $V_{\rm MNS}$ in the same context, but with 
degenerate right-handed neutrino masses, is also consistent with
the experiment.

\end{itemize}

\section{Outlook}

The fact that $V_R$ is very constrained once a hierarchical structure for the Yukawas
is assumed, has an important impact on several physical issues.

\subsection*{Constraints from ${\bf U_{\rm MNS}}$}

We have explored the constraints on $V_R$ from the peculiar pattern of physical
neutrino masses. Similarly, the experimental $U_{\rm MNS}$ may constrain the
high-energy parameters. Although we have seen, eq.(\ref{VLW}), that $V_L$
can always be adjusted to give the observed $U_{\rm MNS}$, it is not guaranteed
that such choice is without tunings for all the possible $V_R$. This may shed
additional light on the structure of the high-energy theory.

\subsection*{Relation to the $R$--parametrization}

The connection of the botton-up parametrization (\ref{param1}), based
on an orthogonal complex matrix $R$ and the top-down parametrization
(\ref{seesaw2}--\ref{VLW}), based on the $V_R$ matrix, is given
in (\ref{YYdag}). Nevertheless, it would be very helpful for 
phenomenological studies to determine from the beginning the form of
$R$ consistent with e.g. hierarchical neutrino Yukawa couplings. This
would give an indication about which $R$s are more natural, and would
make easier in general the exploration of phenomenological signatures
of top-down assumptions.

\subsection*{Leptogenesis}

If one ignores flavour effects, the rate of leptogenesis
produced by the decay of the right-handed neutrinos
is proportional to particular entries of the matrix
\bea
\label{YYlepto}
Y Y^\dagger =  V_R D_{Y^2} V_R^\dagger  \ .
\eea
where $D_{Y^2}= {\rm diag}\{y_1^2,y_2^2,y_3^2\}$.
Since the assumption of hierarchical $y_i$ strongly constrains $V_R$, the
corresponding results for leptogenesis are directly affected.

For the two-neutrino case (see sect. 3), the implications are particularly
nitid: the CP Majorana phase of $V_R$ (the only source of CP violation
for this issue) must be close to a CP-conserving value, which would make
the leptogenesis process inefficient. Nevertheless, flavour effects
can rescue this scenario when the temperature at which leptogenesis
takes place is smaller than $\sim 10^{12}$GeV, as was shown
in \cite{Abada:2006fw} (note that this scenario would correspond to the case $R$ real).
The analysis for three neutrinos is a bit more involved
but it has an obvious interest.

In a supersymmetric framework, another mechanism to generate
the observed baryon asymmetry is Affleck-Dine leptogenesis \cite{Affleck:1984fy}.
Thermal effects and gravitino overproduction constrain
the smallest neutrino mass to be $m_1\lsim 10^{-8}$eV \cite{Asaka:2000nb}.
Despite the large hierarchy between $m_2$ and $m_1$ might
seem a priory unnaturally strong, we have shown that it
is in fact a prediction of the see-saw mechanism with the
suggestive ansatz proposed in section 5 [see eq.(\ref{m1tiny})].

\subsection*{Rare LFV processes}

In the context of SUSY, it is well known that even starting with universal soft masses
at high energy, one ends up with flavour-violating entries in the mass-matrices, 
mainly due to the effect of the neutrino Yukawa couplings in the running between 
the high-energy scale ($M_p$ in the gravity-mediated case) and the scale of 
the right-handed masses \cite{Borzumati:1986qx}. Such effect is proportional to
\bea
\label{YYLFV}
Y^\dagger Y =  V_L D_{Y^2} V_L^\dagger  \ .
\eea
Although $V_L$ is not directly constrained from the low-energy spectrum, once
$V_R$ is determined, $V_L$ is obtained from eq.(\ref{VLW}). The corresponding
rates for LFV processes, such as $\mu\rightarrow e, \gamma$, may constrain
further the scenario and offer predictions for present and future experiments.

\subsection*{GUT constructions}

As mentioned above, identifying $V_R$ with $V_{\rm CKM}$ and taking a hierarchy
of neutrino Yukawa couplings (and right-handed masses) equal to that
of the $u-$quarks, is (non-trivially) consistent with the experiment.
It would be very interesting to build a GUT model able to accommodate
this appealing feature.

\subsection*{Anarchic neutrinos}

As mentioned at the end of subsect. 4.1, the
basis-independent top-down parametrization of the see-saw mechanism that we have
used is likely very appropriate to study scenarios of anarchic neutrinos
\cite{Hall:1999sn},
since these are based on statistical
considerations about the
high-energy parameters that define the theory, and it is highly
desirable that these parameters are basis-independent.
We gave there a simple example of how such analysis can be, but clearly
much work could be done in this direction.

\vspace{0.2cm}
\noindent
------------------

\vspace{0.2cm}
\noindent
Work along the above lines is currently in progress.

\section*{Acknowledgements}

We thank C. Savoy for useful discussions.
This work was supported by the Spanish Ministry of Education
and Science through a MEC project (FPA 2004-02015) and by a 
Comunidad de Madrid project (HEPHACOS; P-ESP-00346).
FJ acknowledges the finantial support of the FPU (MEC) grant,
ref. AP-2004-2949.

\section*{Appendix}

Here we summarize some useful formulas concerning
the eigenvalues of a (general or not) matrix.

According to the Gershgorin Circle Theorem, every eigenvalue
of any complex $n\times n$ matrix $A$ lies within at least
one of the $n$ Gershgorin discs defined as
\bea
\label{Discs}
D(A_{ii}, R_i) \equiv \{z: |z-A_{ii}|\leq R_i\} \ .
\eea
where $R_i$ is the Gershgorin radius of the Gershgorin disc centered 
at $A_{ii}$,
\bea
\label{Radius}
R_i = \sum_{j\neq i} \left|A_{ij}\right|\ .
\eea
For the proof, let $\lambda$ be an eigenvalue of $A$ with eigenvector 
${\bf v}\equiv \{v_j\}$. Define $|v_i|={\rm max}_j |v_j|$
(always $|v_i|>0$). Then the eigenvalue equation $A {\bf v}=\lambda{\bf v}$
can be written as
\bea
\label{eigeneq}
\lambda v_i - A_{ii}v_i = 
\sum_{j\neq i} A_{ij}v_j
\eea
Dividing both sides by $v_i$ and taking the norm we obtain
\bea
\label{eigeneq2}
|\lambda - A_{ii}| = \left|
\sum_{j\neq i} A_{ij}v_j/v_i\right|\leq \sum_{j\neq i} |A_{ij}| = R_i
\eea
Working with $A^T$ instead of $A$ we get an analogous expression for the
same eigenvalues changing $R_i \rightarrow \sum_{j\neq i} |A_{ji}|$. I.e.
for each diagonal element, there is one Gershgorin radius associated with 
the row and one with the column. Furthermore it can be shown that if 
the $n$ discs can be partitioned into disjoint subsets of the complex plane then each subset contains the same number of  eigenvalues as discs.

If the original matrix $A$ is hermitian, then the eigenvalues of $A$, 
say $\lambda_i$,
and diagonal elements, $A_{ii}$, are real, so the discs become
segments in the real line. Furthermore, the Gershgorin segments
associated with the rows and the columns coincide.

All this can be applied to eq.(\ref{seesaw3}). In particular, for the case
of three neutrinos with hierarchical Yukawa couplings, $y_1\ll y_2\ll y_3$, 
the diagonal entry $\left(\kappa'\kappa'^\dagger\right)_{33}$ is normally 
much larger than the others and the corresponding Gershgorin radius is much
smaller (see below), so the Gershgorin disc is usually disjoint from the
others. This means that the largest eigenvalue, $\kappa_3^2$, satisfies
\bea
\label{appineq3}
\left|\kappa_3^2 - \left(\kappa'\kappa'^\dagger\right)_{33}\right| 
\leq
\sum_{j\neq 3} \left|\left(\kappa'\kappa'^\dagger\right)_{3j}\right| 
\eea
which is similar to eq.(\ref{3ineq2}) [note that the right-hand-side of
eq.(\ref{appineq3}), i.e. the Gershgorin radius, is supressed by a ${y_2\over y_3}-$factor with respect to 
$\left(\kappa'\kappa'^\dagger\right)_{33}$].

Analogous inequalities can be produced for $\kappa_1^2$. In this case, the most efficient ones come from considering the inverse matrix, $\kappa'^{-1}(\kappa'^{-1})^\dagger$, which is a positive hermitian matrix
with $\kappa_i^{-2}$ eigenvalues.

The inequalities for $\kappa_3$, $\kappa_1$ produced in this way can be
plugged into (\ref{3hie}) to give bounds on $h$ similar to those considered
in sect.~4. 

Let us recall that in that section we found more efficient for the sake of clarity
to use the fact that in a positive hermitian matrix, such as $\kappa'\kappa'^\dagger$
and $\kappa'^{-1}(\kappa'^{-1})^\dagger$, the largest eigenvalue ($\kappa_3^2$ and
$\kappa_1^{-2}$ respectively) must be larger than any diagonal entry of the matrix.

For the proof, let $A$ be a positive hermitian $n\times n$ matrix with 
eigenvalues $\{\lambda_i\}$ and eigenvectors $\{{\bf v_i}\}$, ordered
as $\lambda_1\leq \lambda_2\leq\cdots \lambda_n$. Writing
the normalized vector in the $i^{\rm th}$-direction, ${\bf e_i}$,
as ${\bf e_i}= \sum a_{ij} {\bf v_j} $ with $\sum |a_{ij}|^2=1$, then
\bea
\label{appineq4}
{\bf e_i}^\dagger A {\bf e_i} = A_{ii}=
\sum \lambda_j |a_{ij}|^2 \leq \lambda_n
\eea
Similarly it can be shown that $\lambda_1\leq A_{ii}$ (for any $i$).
The above lower bound for $\lambda_n$ is complemented with the obvious
upper bound $\lambda_n \leq {\rm tr}A$. This allows to corner the
range of values where $\lambda_n$ lies. (This  procedure is very efficient
for $\kappa'\kappa'^\dagger$ and $\kappa'^{-1}(\kappa'^{-1})^\dagger$,
since the trace is strongly dominated by the largest diagonal entry.)

These inequalities can be made stronger replacing $A_{ii}$ by the 
eigenvalues of any $m\times m$ submatrix of $A$ (with $m\leq n$). This can be seen by
diagonalizing the submatrix with $A'=V^\dagger A V$, where $V$ is a unitary matrix which is trivial except in the corresponding
$m\times m$ box, and then applying the same argument to $A'$.

All  these kinds of inequalities can be plugged into (\ref{3hie})
to obtain alternative bounds on $h$. Also they can be used to 
put bounds on the ratio $\kappa_3/\kappa_1$, which gives a direct measure
of how far is the neutrino spectrum from the exactly degenerate case.


\begin{thebibliography}{99}


\bibitem{Gonzalez-Garcia:2002dz}
  For a review on neutrino physics, see 
 M.~C.~Gonzalez-Garcia and Y.~Nir,
  Rev.\ Mod.\ Phys.\  {\bf 75} (2003) 345
  [arXiv:hep-ph/0202058].
%
\bibitem{Maltoni:2004ei}
  M.~Maltoni, T.~Schwetz, M.~A.~Tortola and J.~W.~F.~Valle,
  New J.\ Phys.\  {\bf 6} (2004) 122
  [arXiv:hep-ph/0405172]. (Updated June--2006)
%
\bibitem{Yao:2006px}
  W.~M.~Yao {\it et al.}  [Particle Data Group],
  J.\ Phys.\ G {\bf 33} (2006) 1.
%
\bibitem{seesaw} 
P.~Minkowski,
Phys.\ Lett.\ B {\bf 67} (1977) 421.
M. Gell-Mann, P. Ramond and R. Slansky, \emph{Proceedings
of the Supergravity Stony Brook Workshop}, New York 1979, eds. P. Van
Nieuwenhuizen and D. Freedman; 
T. Yanagida, \emph{Proceedinds of the
Workshop on Unified Theories and Baryon Number in the Universe}, Tsukuba,
Japan 1979, eds. A. Sawada and A. Sugamoto; 
R. N. Mohapatra, G. Senjanovic, 
\textit{Phys.Rev.Lett.} \textbf{44} (1980)912, \textit{ibid.} \textit{%
Phys.Rev.} \textbf{D23} (1981) 165; 
S.~L.~Glashow, \emph{The Future Of Elementary Particle Physics},
\textit{In *Cargese 1979, Proceedings, Quarks and Leptons*, 687-713 and
Harvard Univ.Cambridge - HUTP-79-A059 (79,REC.DEC.) 40p}.
%

\bibitem{Pascoli:2003uh}
  S.~Pascoli, S.~T.~Petcov and W.~Rodejohann,
LFV
  Phys.\ Rev.\ D {\bf 68} (2003) 093007
  [arXiv:hep-ph/0302054].
\bibitem{Casas:2001sr}
  J.~A.~Casas and A.~Ibarra,
  Nucl.\ Phys.\ B {\bf 618} (2001) 171
  [arXiv:hep-ph/0103065].


\bibitem{Broncano:2003fq}
  A.~Broncano, M.~B.~Gavela and E.~Jenkins,
  Nucl.\ Phys.\ B {\bf 672} (2003) 163
  [arXiv:hep-ph/0307058].;
  S.~F.~King,
  arXiv:hep-ph/0610239;
  S.~Antusch and S.~F.~King,
  JHEP {\bf 0601} (2006) 117
  [arXiv:hep-ph/0507333].

\bibitem{Frampton:2002qc}
  P.~H.~Frampton, S.~L.~Glashow and T.~Yanagida,
  Phys.\ Lett.\ B {\bf 548} (2002) 119
  [arXiv:hep-ph/0208157];
  M.~Raidal and A.~Strumia,
  Phys.\ Lett.\ B {\bf 553} (2003) 72
  [arXiv:hep-ph/0210021];
  A.~Ibarra and G.~G.~Ross,
  Phys.\ Lett.\ B {\bf 591} (2004) 285
  [arXiv:hep-ph/0312138];
  W.~l.~Guo, Z.~z.~Xing and S.~Zhou,
  arXiv:hep-ph/0612033.


\bibitem{Hall:1999sn}
  L.~J.~Hall, H.~Murayama and N.~Weiner,
  Phys.\ Rev.\ Lett.\  {\bf 84} (2000) 2572
  [arXiv:hep-ph/9911341];
  N.~Haba and H.~Murayama,
  Phys.\ Rev.\ D {\bf 63} (2001) 053010
  [arXiv:hep-ph/0009174];
  A.~de Gouvea and H.~Murayama,
  Phys.\ Lett.\ B {\bf 573} (2003) 94
  [arXiv:hep-ph/0301050];
  J.~R.~Espinosa,
  arXiv:hep-ph/0306019.



\bibitem{Ramond}
  L.~Everett and P.~Ramond,
  arXiv:hep-ph/0608069.

\bibitem{Casas:1999tg}
  J.~A.~Casas, J.~R.~Espinosa, A.~Ibarra and I.~Navarro,
  Nucl.\ Phys.\ B {\bf 573} (2000) 652
  [arXiv:hep-ph/9910420].

\bibitem{Battaglia:2003in}
See, for instance, M.~Battaglia {\it et al.},
``The CKM matrix and the unitarity triangle,''
[hep-ph/0304132].

\bibitem{Olechowski:1990bh}
  M.~Olechowski and S.~Pokorski,
  Phys.\ Lett.\ B {\bf 257} (1991) 388.

\bibitem{Ramond:1993kv}
  P.~Ramond, R.~G.~Roberts and G.~G.~Ross,
  Nucl.\ Phys.\ B {\bf 406} (1993) 19
  [arXiv:hep-ph/9303320].

\bibitem{Abada:2006fw}
  A.~Abada, S.~Davidson, F.~X.~Josse-Michaux, M.~Losada and A.~Riotto,
  JCAP {\bf 0604}, 004 (2006)
  [arXiv:hep-ph/0601083];
  E.~Nardi, Y.~Nir, E.~Roulet and J.~Racker,
  JHEP {\bf 0601}, 164 (2006)
  [arXiv:hep-ph/0601084];
  A.~Abada, S.~Davidson, A.~Ibarra, F.~X.~Josse-Michaux, M.~Losada and
A.~Riotto,
  JHEP {\bf 0609}, 010 (2006)
  [arXiv:hep-ph/0605281].


\bibitem{Affleck:1984fy}
  I.~Affleck and M.~Dine,
  Nucl.\ Phys.\ B {\bf 249}, 361 (1985).


\bibitem{Asaka:2000nb}
  T.~Asaka, M.~Fujii, K.~Hamaguchi and T.~Yanagida,
  Phys.\ Rev.\ D {\bf 62}, 123514 (2000)
  [arXiv:hep-ph/0008041].

\bibitem{Borzumati:1986qx}
  F.~Borzumati and A.~Masiero,
  Phys.\ Rev.\ Lett.\  {\bf 57}, 961 (1986).

\end{thebibliography}
\end{document}